\documentclass[onecolumn,notitlepage,nofootinbib,superscriptaddress]{revtex4-1}
\pdfoutput=1
\usepackage{amsmath}
\usepackage{array}
\usepackage{graphicx}
\usepackage{multirow}
\usepackage{pbox}
\usepackage[position=t,singlelinecheck=off,justification=raggedright]{subfig}

\begin{document}




\title{ Raman Spectroscopy of Graphene }


\author{ Sven Reichardt }
\affiliation{ Physics and Materials Science Research Unit, Universit\'e du Luxembourg, 1511 Luxembourg, Luxembourg }
\affiliation{ JARA-FIT and 2nd Institute of Physics, RWTH Aachen University, 52074 Aachen, Germany }

\author{ Ludger Wirtz }
\affiliation{ Physics and Materials Science Research Unit, Universit\'e du Luxembourg, 1511 Luxembourg, Luxembourg }
             



\begin{abstract}

Raman spectroscopy of graphene is reviewed from a theoretical perspective.
After an introduction of the building blocks (electronic band structure, phonon dispersion, electron-phonon interaction, electron-light coupling), Raman intensities are calculated using time-dependent perturbation theory.
The analysis of the contributing terms allows for an intuitive understanding of the Raman peak positions and intensities.
The Raman spectrum of pure graphene only displays two principle peaks.
Yet, their variation as a function of internal and external parameters and the occurrence of secondary, defect-related peaks, conveys a lot of information about the system.
Thus, Raman spectroscopy is used routinely to analyze layer number, defects, doping and strain of graphene samples.
At the same time, it is an intriguing playground to study the optical properties of graphene.

\end{abstract}


\maketitle
\tableofcontents



\section{Introduction}

Raman scattering is the inelastic scattering of light.
The frequency of a photon can change by transferring energy to and/or receiving energy from the lattice vibrations of the material.
In quantum mechanics, the ``allowed'' energies of an oscillation are quantized and - in the case of a harmonic oscillation - equidistant.
The vibrations of the lattice can thus be described in the language of quasi-particles and the term \emph{phonon} is usually used.
The incoming photon changes its energy by effectively exciting or absorbing one or several phonons.
If a phonon is excited, the photon looses energy and the process is called \emph{Stokes} scattering.
In the case of the absorption of a phonon, the photon gains energy and the process is referred to as \emph{Anti-Stokes} scattering.
The spectrum of the inelastically scattered light therefore features discrete peaks, whose positions can be directly associated with certain vibrational modes of the crystal.

In first order Raman scattering, only optical phonons at the Brillouin zone center can be excited.
Furthermore, selection rules only allow for vibrations with special symmetry properties.
For graphene, only one peak, the so-called $G$~peak, is due to first-order Raman scattering.
An additional line, the so-called $2D$~line, is due to second order Raman scattering, where two phonons are excited at the same time.
The overall Raman spectrum of graphene thus consists mainly of two peaks (see Fig.~\ref{fig:example-spectrum}) and is quite similar to the spectrum of graphite and of carbon nanotubes.
Nevertheless, via the exact peak positions, their width, their dispersion as a function of the exciting laser energy, and the occurrence of side peaks, Raman spectroscopy yields a surprising amount of information about graphene samples:
For example, it gives information on the number of layers, their stacking order, the underlying substrate, defects, impurities, doping, and strain.
Since Raman scattering is a fast and non-invasive method, it has evolved into one of the principal characterization tools for graphene and related graphitic materials.

Raman spectroscopy is often categorized as ``vibrational spectroscopy'' because it probes primarily the phonon frequencies.
However, a quantitative description of the Raman intensities is deeply connected to the optical (absorption and emission) properties of the concerned material.
This is particularly the case for resonant Raman spectroscopy, where the emitted/absorbed light is in resonance with electronic transitions.
In graphene, due to the linear crossing of the optically active $\pi$ and $\pi^*$ bands, the Raman effect is resonant for all laser frequencies in the infrared, visible, and near UV ranges.
Different ``quantum pathways''~\cite{chen2011} (corresponding to contributions of different microscopic processes in perturbation theory) contribute to the intensity of a given peak.
This gives rise to intriguing quantum interference effects with sometimes surprising consequences for the Raman intensities (e.g., as a function of laser energy and of sample doping).
Furthermore, the intensities and the widths of the Raman peak are strongly influenced by the lifetimes of intermediate electronic excitations.
Raman scattering thus also bears information on the complex dynamics of electron-hole pairs in graphene.
All this makes a chapter on Raman spectroscopy an integral part of a book about the optical properties of graphene.

A number of books/book chapters~\cite{saito1998,reich2004} and review articles~\cite{ferrari2007,malard2009,ferrari2013} have already been devoted to the topic of Raman scattering in graphitic materials.
This book chapter puts particular emphasis on the theoretical foundations of resonant Raman scattering in graphene.
It explains the basic theory for Raman scattering in graphene using perturbation theory.
Together with the study of the groundbreaking theoretical research articles (in particular~\cite{basko2008,basko2009,venezuela2011,herziger2014}), it should enable the newcomer to the field to learn how an exact quantitative calculation can be achieved in principle.
For the computationally less ambitious reader, the chapter gives detailed qualitative understanding how the different terms in perturbation theory depend on external stimuli (such as doping, defects, or strain) and what conclusions one can draw from the Raman spectra.

The chapter is structured as follows:
In section~\ref{sec:theory}, we first introduce the basic building blocks for the understanding of resonant Raman spectra in graphene:
The electronic band structure, the phonon dispersion, the coupling between electrons and phonons, and the coupling of the electronic states to light.
Afterwards, the formalism of time-dependent perturbation theory is applied to give the reader an intuitive understanding and enable her/him to study the relevant research literature.
In section~\ref{sec:applications}, the theory is then applied to understand qualitatively and quantitatively, how Raman spectroscopy can be used to study the quality of graphene flakes as well as to understand the effects of environment, doping and strain.

\section{Theoretical description of the Raman spectrum of graphene}
\label{sec:theory}

In this first section, we will have a look at the microscopic origin of the Raman spectrum of graphene, in particular the behavior, shape, and size of its various peaks.
To this end, we will use time-dependent perturbation theory to calculate expressions for the Raman scattering amplitudes.
This approach has the advantage that the individual terms of the perturbation series can be represented by so-called \emph{Feynman diagrams}.
The graphical representation of the calculation allows for both easier bookkeeping of the terms in the perturbation series and at the same time gives a simple physical picture for the microscopic processes governing Raman scattering.

In the first part of this section, we will introduce the basic building blocks needed for the perturbation theory calculation.
We briefly discuss the electronic band structure of graphene before having a more detailed look at the phonon dispersion of graphene.
Thereafter, we will discuss the coupling between electrons and phonons and their treatment in perturbation theory and show how to include the effects of an external electromagnetic field, i.e.  of incoming and outgoing photons.
This first part will be concluded by briefly touching upon electron-defect scattering and a summary in which we give a graphical representation of the  mathematical expressions for the various building blocks in terms of Feynman diagrams.

The second part contains an overview over the kinematics of Raman scattering.
We will derive what kind of peaks are expected in the Raman spectrum of graphene based on kinematical considerations alone.

In the third subsection, we will focus on the actual calculation of the Raman spectrum.
Using time-dependent perturbation theory and the language of Feynman diagrams, we will demonstrate how the Raman spectrum of graphene can be calculated.
in this section we will mainly focus on the two most prominent Raman peaks and study their behavior from the scattering amplitude obtained in perturbation theory.

\subsection{Basic building blocks}

We start by introducing the basic ingredients needed to theoretically describe the Raman spectrum of graphene.
These ingredients are the electronic band structure, the phonon dispersion, the electron-phonon coupling, the electron-light coupling, and (if the sample is not perfect) electron-defect scattering.
With the exception of the latter, we will have a look at each of these building blocks before we conclude this section with a summary.

\subsubsection{Electronic band structure}

\begin{figure}[htb]
	\centering
	\includegraphics[scale=0.25]{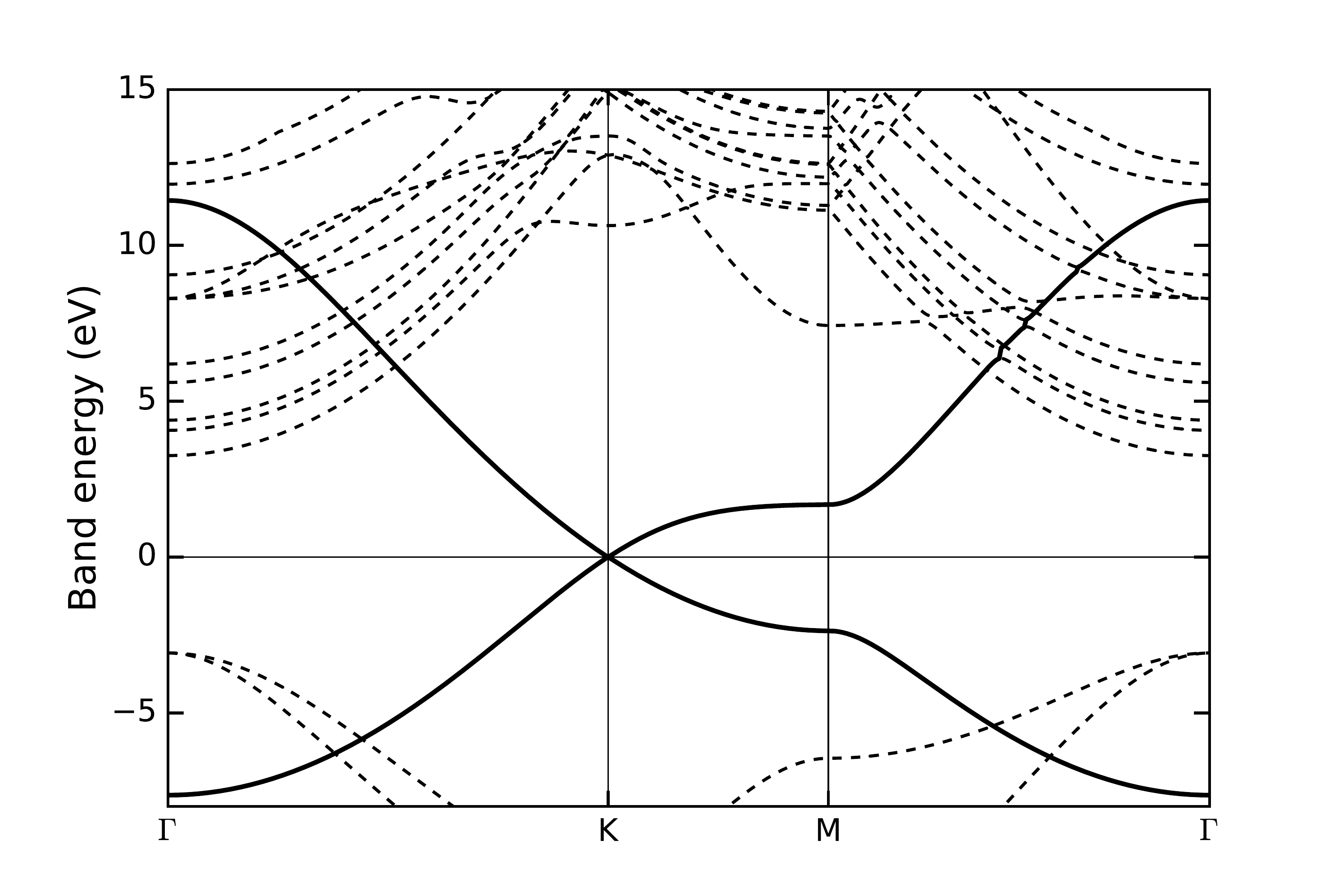}
	\caption{
	{\bf Electronic band structure of monolayer graphene.}
	{\it Ab initio} electronic band structure of monolayer graphene along the high symmetry line $\Gamma - K - M - \Gamma$ obtained from density functional theory (DFT) in the generalized gradient approximation (GGA).
	The full lines are the $\pi$ and $\pi^*$ bands that are by far the most important bands for the Raman scattering process.
	}
	\label{fig:lat-elec-bands}
\end{figure}

Graphene is a monolayer of carbon atoms arranged in a two-dimensional hexagonal lattice.
Each carbon atom possesses four valence electrons, three of which occupy hybridized $sp^2$ orbitals, which form strong, in-plane covalent $\sigma-$bonds.
The forth valence electron occupies a $p_z$-orbital, which form $\pi$-bonds, in which the electrons are delocalized.
The electronic band structure, obtained from an {\it ab initio} calculation using density functional theory (DFT) is shown in Fig.~\ref{fig:lat-elec-bands}.

Since we will be interested in describing and calculating the Raman spectrum for excitation energies in or near the visible spectral range ($\sim$1.5-4~eV), it is sufficient to focus on the bands near the Fermi energy, which are the $\pi$ and $\pi^*$ bands (full lines in Fig.~\ref{fig:lat-elec-bands}).
Note that for a quantitatively accurate calculation of the Raman spectrum, the DFT band structure needs to be corrected to properly include electron-electron-interaction effects.
These corrections can be computed, for example, within the $GW$~approximation~\cite{hedin1965,onida2002} and their main effect is a steepening of the conic $\pi$~bands around the $K$~point with their slope increasing by roughly 18\%~\cite{gruneis2008,venezuela2011}.

\subsubsection{Phonon dispersion}

\begin{figure}[htb]
	\centering
	\subfloat[]{
		\centering
		\includegraphics[scale=0.25]{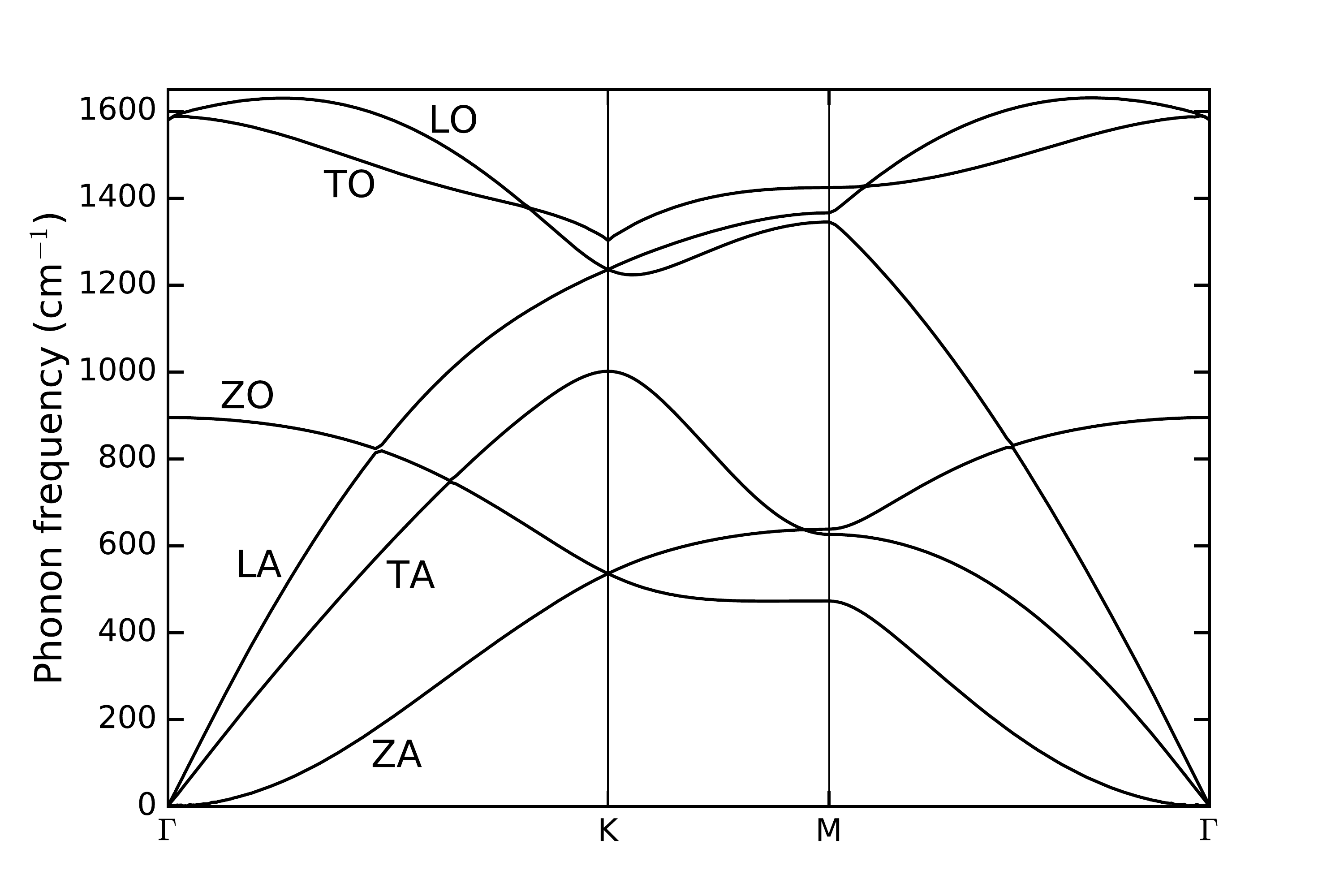}
	}
	\subfloat[]{
		\centering
		\includegraphics[scale=0.18]{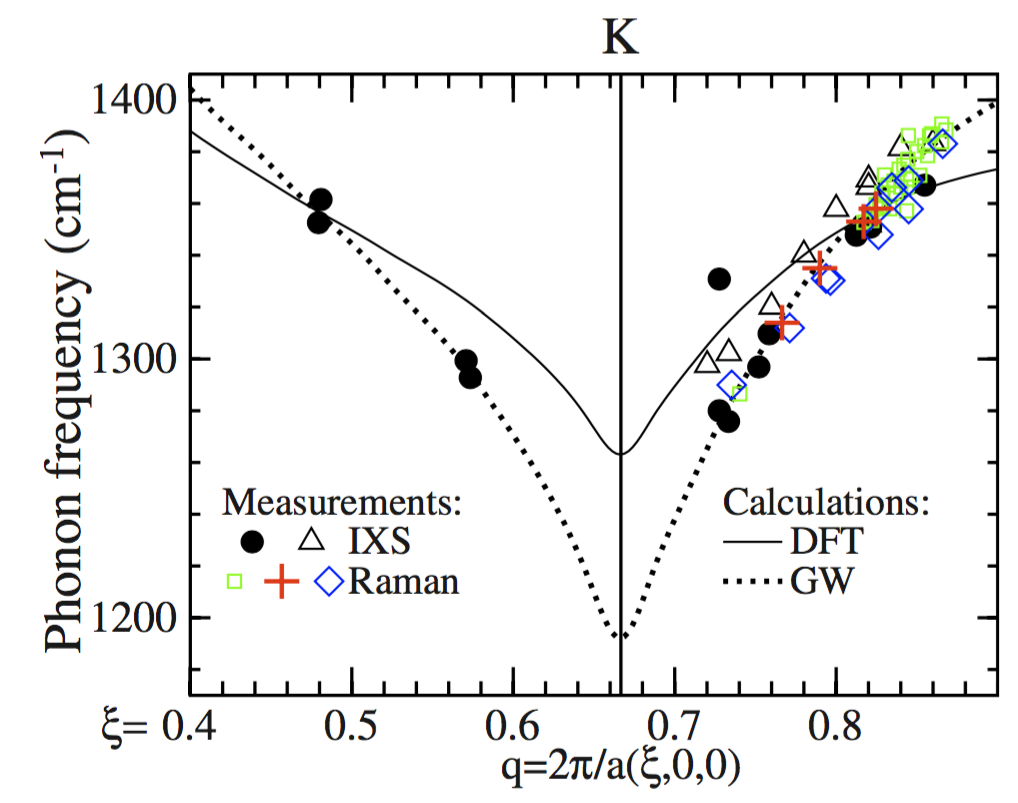}
	}
	\caption{
	{\bf Phonon dispersion of monolayer graphene.}
	(a) {\it Ab initio} phonon dispersion of monolayer graphene along the high symmetry line $\Gamma - K - M - \Gamma$ obtained from density functional perturbation theory (DFPT) in the local density approximation (LDA).
	(b) Zoom-in into the phonon dispersion for phonon wave vectors around $K$.
	Symbols denote experimental data obtained from inelastic X-ray scattering (IXS) and Raman scattering.
	Full lines represent the TO branch obtained from density functional theory within the generalized gradient approximation (GGA).
	Dashed lines represents TO phonon frequencies corrected within perturbation theory, where the electron-phonon coupling was obtained on the level of the $GW$~approximation.
    (Figure taken from Ref.~\onlinecite{lazzeri2008}.)
	}
	\label{fig:phon-dispersion}
\end{figure}

Besides the electronic band structure, the phonon dispersion of graphene also plays a very important role for the understanding of the Raman spectrum.
Since the lattice structure of graphene can be described by a hexagonal lattice with two atoms per unit cell, the phonon band structure consists of 2$\times$3=6 branches:
Three acoustic ones, which can be further divided into an in-plane longitudinal (LA), an in-plane transverse (TA) and an out-of-plane transverse (ZA) branch, and three optical branches, which can also be separated into in-plane longitudinal (LO), in-plane transverse (TO), and out-of-plane transverse (ZO).
The terms ``longitudinal'' and ``transverse'' refer to displacements in directions parallel and perpendicular to the quasimomentum $\mathbf{q}$ of the phonon, respectively.
Note that these notions only make sense for small phonon wave vectors, i.e for $\mathbf{q}$ near $\Gamma$.
Away from $\Gamma$, the character of the modes is a mixture of acoustic (in-phase oscillations of neighboring atoms) and optical (opposite phase oscillations of neighboring atoms) behavior.
However, for simplicity the labels LA etc. are usually also used to designate the phonon branches away from $\Gamma$ by following the specific branch along the high-symmetry lines.
The phonon dispersion of graphene as obtained from a density functional perturbation theory (DFPT) calculation is shown in Fig.~\ref{fig:phon-dispersion}a.

\begin{figure}[htb]
	\centering
	\subfloat[]{
		\centering
		\includegraphics[scale=0.4]{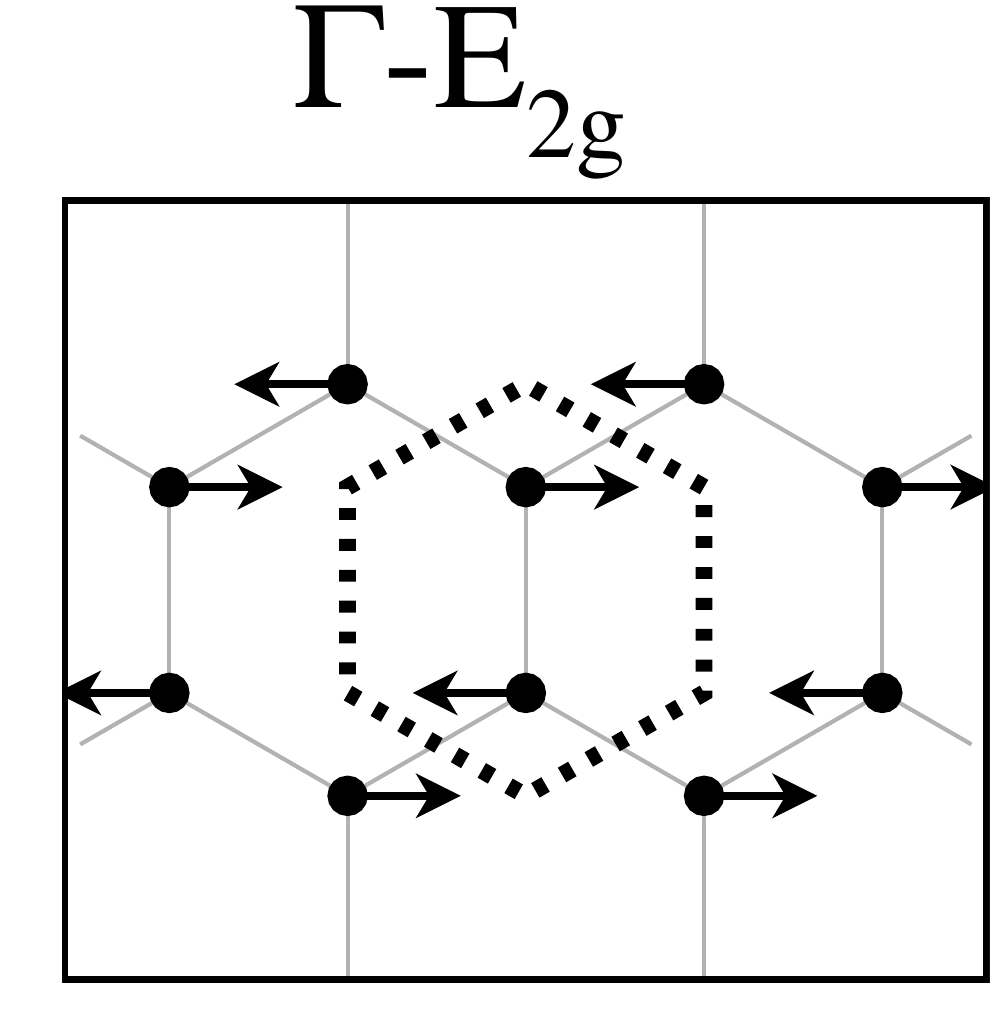}
	}
	\subfloat[]{
		\centering
		\includegraphics[scale=0.4]{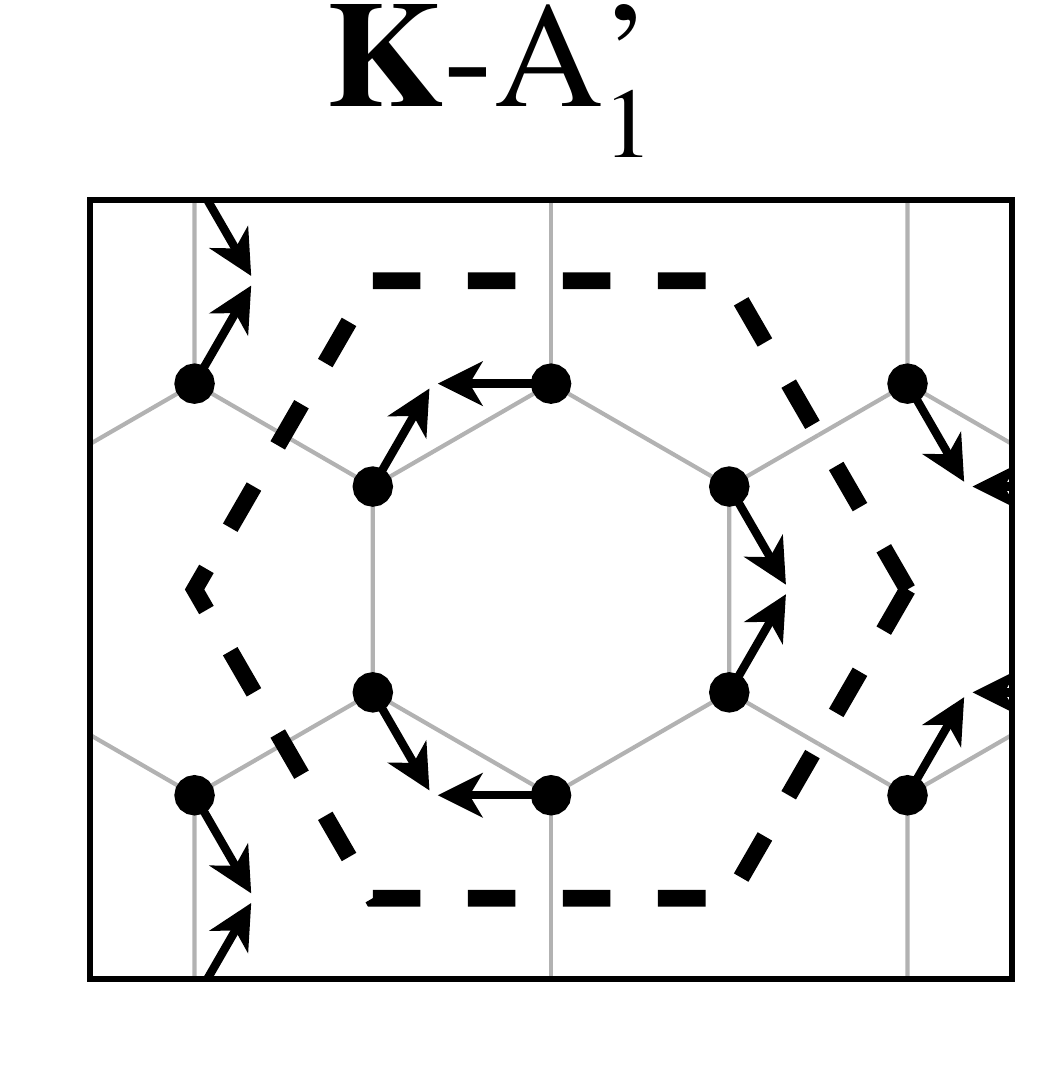}
	}
	\caption{
	{\bf Lattice vibration patterns.}
	(a) Vibration pattern corresponding to one of the two degenerate, optical in-plane phonons at $\Gamma$.
	In the commonly used \emph{Mulliken notation}, this mode is called an $E_{2g}$~mode, as it transforms in the $E_{2g}$~representation of the point group of $\Gamma$.
	The dotted line marks the boundaries of the Wigner-Seitz cell of the graphene lattice.
	(b) Vibration pattern corresponding to the TO branch at $K$.
	In the Mulliken notation, this mode is denoted as $A'_1$, as it transforms in the $A'_1$~representation of the point group of $K$.
	The dashed lines denote the boundaries of a supercell of six atoms.
    (Figure taken from Ref.~\onlinecite{lazzeri2008}.)
	}
	\label{fig:displacements}
\end{figure}

The most prominent features of the phonon dispersion are the kink in the TO~branch at $K$ and the positive slope of the LO~branch at $\Gamma$.
As we will see in a later section, these are precisely the parts of the phonon dispersion that play the most important role in the description of the Raman spectrum as the phonons that contribute the most to it are the two degenerate in-plane optical phonons at $\Gamma$ and the phonons of the TO branch around the $K$~point, whose lattice vibration patterns are shown in Fig.~\ref{fig:displacements}.
For a description of the Raman spectrum it is therefore of utmost importance to accurately describe these parts of the phonon dispersion.

The pronounced kinks arise due to so-called \emph{Kohn anomalies}~\cite{kohn1959}.
When the lattice ions vibrate according to a phonon mode with wave vector $\mathbf{q}$, they induce an electronic charge density of the same periodicity, whose magnitude becomes large when $\mathbf{q}$ connects two points on the Fermi surface.
Since the Fermi surface of graphene consists of the two points $K$ and $K'$ at the corners of the Brillouin zone (which are separated by a vector $\mathbf{K}$), phonons with $\mathbf{q} = \mathbf{0}$ or $\mathbf{q} = \mathbf{K}$ lead to these large, periodic electronic charge densities.
These electronic charge densities in turn lead to descreased screening of the ionic charges, leading to a great enhancement of electron-phonon interaction.
This large electron-phonon coupling for $\mathbf{q}$ near $\Gamma$ or $K$ is responsible for the comparatively steep slope of the phonon dispersion (i.e. the kinks) in these parts of the Brillouin zone, since it can be shown that at $\Gamma$ and $K$, the slope of the phonon dispersion is directly proportional to the squared electron-phonon coupling between the corresponding phonon and electrons in the $\pi$ and $\pi^*$ bands averaged over a small circle of electronic $\mathbf{k}$ vectors around the Fermi points $K$ and $K'$~\cite{piscanec2004}.

The increased electron-phonon coupling at $\Gamma$ and especially at $K$ is further strongly enhanced by electronic correlation effects~\cite{lazzeri2008}.
This manifests itself in the inaccurate quantitative description of the phonon dispersion near $K$ in density functional theory, where correlation effects due to electron-electron interaction are not well-described by exchange-correlation functionals in the commonly used local density or generalized gradient approximations.
As will be commented upon in the next subsection, it is possible to obtain a more accurate description of the electron-phonon coupling for phonons at $\Gamma$ and $K$ on the level of the $GW$~approximation.
These corrected values for the electron-phonon coupling lead to a correction of the phonon frequencies within perturbation theory.
The most important result is a steepening of the kink near $K$ and a decrease of the phonon frequency of the TO~branch at $K$ by more than 100~cm$^{-1}$~\cite{lazzeri2008}, which matches the experimental data well (compare Fig.~\ref{fig:phon-dispersion}b).

\subsubsection{Electron-phonon coupling}

Besides a description of the electron and phonon band structures, one needs a description of the coupling between the two.
For this, we consider the electronic Hamiltonian (discarding electron-electron interaction for the moment), which has the form
\begin{equation}
H_0 = \frac{\mathbf{p}^2}{2m_e} + V(\mathbf{r};\{\mathbf{R}_{n\alpha,0}\}),
\end{equation}
where $\mathbf{r}$ and $\mathbf{p}$ denotes the position and momentum operator, respectively, and $V$ is the lattice potential which depends parametrically on the position of the nuclei, which we collectively denote with $\{\mathbf{R}_{n\alpha}\}$, where $n$ labels the unit cell and $\alpha$ specifies the atom within the unit cell.
The index ``0'' refers to the equilibrium positions of the nuclei.

When the lattice atoms are oscillating, the position of the nucleus at position $\mathbf{R}_{n\alpha,0}$ changes by a small displacement: $\mathbf{R}_{n\alpha,0} \to \mathbf{R}_{n\alpha} = \mathbf{R}_{n\alpha,0} + \mathbf{u}(\mathbf{R}_{n\alpha,0})$, where $\mathbf{u}(\mathbf{R}_{n\alpha,0})$ denotes the displacement of the nucleus at equilibrium position $\mathbf{R}_{n\alpha,0}$.
As a consequence, the lattice potential changes as well and for small displacements $\mathbf{u}(\mathbf{R}_{n\alpha,0})$ the potential can be expanded into a Taylor series:
\begin{equation}
\begin{split}
V(\mathbf{r};\{\mathbf{R}_{n\alpha}\}) &=  V(\mathbf{r};\{\mathbf{R}_{n\alpha,0}\}) + \sum_{n,\alpha} \frac{\partial V}{\partial \mathbf{R}_{n\alpha}} (\mathbf{r};\{\mathbf{R}_{n\alpha,0}\}) \cdot \mathbf{u}(\mathbf{R}_{n\alpha,0}) \\
                       &  \quad + \frac{1}{2} \sum_{n,m,\alpha,\beta} \sum_{i,j} \frac{\partial^2 V}{\partial R_{n\alpha,i} \partial R_{m\beta,j}} (\mathbf{r};\{\mathbf{R}_{n\alpha,0}\}) u_i(\mathbf{R}_{n\alpha,0}) u_j(\mathbf{R}_{m\beta,0}) + \mathcal{O}(u^3).
\end{split}
\end{equation}
Here, the sums run over all unit cells $n$, all atoms $\alpha$ within the unit cell, and all cartesian coordinates $i$.
The electronic Hamiltonian thus acquires two more terms (up to quadratic order in the displacement), describing the coupling of electrons to one and two phonons, respectively:
\begin{equation}
H_0 \to H_0 + H^{(1)}_{\mathrm{el-ph}} + H^{(2)}_{\mathrm{el-ph}},
\end{equation}
where $H^{(i)}_{\mathrm{el-ph}}$ is given by the $i$th order term in the expansion of the potential given above.

After quantizing the phonon field, the operator corresponding to a displacement of the atom at equilibrium position $\mathbf{R}_{n\alpha,0}$ can be written as a superposition of the vibrational eigenmodes:
\begin{equation}
\mathbf{u}(\mathbf{R}_{n\alpha,0}) = \sum_{\mathbf{q},\lambda} \sqrt{\frac{\hbar}{2 M_C \omega_{\mathbf{q},\lambda}}} \left[ b_{\lambda,\mathbf{q}} \mathbf{v}^{\alpha}_{\lambda,\mathbf{q}} + b^{\dagger}_{\lambda,-\mathbf{q}} \left(\mathbf{v}^{\alpha}_{\lambda,-\mathbf{q}}\right)^* \right] \mathrm{e}^{i \mathbf{q} \cdot \mathbf{R}_n},
\end{equation}
where $\mathbf{q}$ and $\lambda$ denote the quasimomentum and phonon branch (e.g. $\lambda$=TO for the transverse optical branch) of the eigenmode, respectively, $\mathbf{R}_n$ is the position, i.e., lattice site, of the $n$th unit cell, and $b^{(\dagger)}_{\lambda,\mathbf{q}}$ is an operator that destroys (creates) one phonon of branch $\lambda$ with quasimomentum $\mathbf{q}$.
The vector $\mathbf{v}^{\alpha}_{\lambda,\mathbf{q}}$ denotes one of the two three-dimensional parts of the six-dimensional eigenvector of the dynamical matrix describing the displacement of atom $\alpha$ of the unit cell.

To describe the Raman scattering amplitude within perturbation theory, one needs the matrix elements of the electron-phonon Hamiltonian for the $\pi$ and $\pi^*$ states and for one or two specific phonons.
For example, the matrix elements of the electron-one-phonon Hamiltonian for a phonon of branch $\lambda$ and with quasimomentum $\mathbf{q}$ in the final state are given by
\begin{equation}
\begin{split}
g^{\lambda}_{s,\mathbf{k}-\mathbf{q};s',\mathbf{k}} &= \langle s,\mathbf{k}-\mathbf{q}; \lambda,\mathbf{q} | H^{(1)}_{\mathrm{el-ph}} | s', \mathbf{k} \rangle \\
                                           &=  \sqrt{\frac{\hbar}{2 M_C \omega_{\mathbf{q},\lambda}}} \sum_{n,\alpha} \mathrm{e}^{i \mathbf{q} \cdot \mathbf{R}_n}
                                               \left(\mathbf{v}^{\alpha}_{\lambda,\mathbf{q}}\right)^* \\
                                           & \quad \quad \cdot \bigg\langle s,\mathbf{k}-\mathbf{q} \bigg| \frac{\partial V}{\partial \mathbf{R}_{n\alpha}} (\mathbf{r};\{\mathbf{R}_{n\alpha,0}\}) \bigg| s',\mathbf{k} \bigg\rangle,
\end{split}
\end{equation}
where $s$ and $s'$ specify the electronic band and $|s,\mathbf{k}-\mathbf{q}; \lambda,\mathbf{q}\rangle$ is a state consisting of an unperturbed electronic state specified by the band index $s$ and Bloch wave vector $\mathbf{k}-\mathbf{q}$ and a phonon from branch $\lambda$ with quasimomentum $\mathbf{q}$.
For the Raman spectrum of graphene, only the $\pi$ and $\pi^*$ bands are important, i.e. $s,s' = \pi,\pi^*$ and for every $\mathbf{k}$, $\mathbf{q}$, and $\lambda$ the electron-phonon coupling matrix elements can be written as a 2$\times$2-matrix $\mathbf{g}^{\lambda}_{\mathbf{k}-\mathbf{q},\mathbf{q}} = \left(g^{\lambda}_{s,\mathbf{k}-\mathbf{q};s',\mathbf{k}}\right)_{s,s'=\pi,\pi^*}$ in the space of the $\pi$ and $\pi^*$ bands.\\
\\
By the same procedure, one can obtain a 2$\times$2-matrix containing the matrix elements of the electron-two-phonon interaction Hamiltonian $H^{(2)}_{\mathrm{el-ph}}$.
For the purpose of this chapter, only those matrix elements are important that contain the two phonons in the final state.
Again, these can conveniently written as a 2$\times$2 basis in the space of $\pi$ and $\pi^*$ bands, which we will denote by $\mathbf{\tilde{g}}^{\lambda_1,\lambda_2}_{\mathbf{k}-\mathbf{q}_1-\mathbf{q}_2,\mathbf{k}}$, where $\lambda_{1(2)}$ and $\mathbf{q}_{1(2)}$ are the quantum numbers of phonon 1(2).\\
\\
Numerical values for the electron-phonon matrix elements can be obtained from density functional perturbation theory.
It should be noted, however, that the underestimation of electron-electron interaction effects in DFT leads to an underestimation of the electron-phonon matrix elements as well, especially for $\mathbf{k}$ and $\mathbf{q}$ near $K$, where long-range electronic correlation effects play an important role.
For specific $\mathbf{k}$ and $\mathbf{q}$, such as $\mathbf{k}=\mathbf{q}=K$, it is possible to incorporate these effects by relating the electron-phonon matrix elements at these high-symmetry points to changes of the band energies when the lattice ions are displaced statically according to the phonon displacement pattern shown in Fig.~\ref{fig:displacements}b.
For instance, for a phonon from the TO branch with quasimomentum $\mathbf{q}=\mathbf{K}$, the square of the electron-phonon coupling for a $\pi$ and $\pi^*$ state with $\mathbf{k} = 2\mathbf{K} = \mathbf{K}'$ obeys~\cite{piscanec2004,lazzeri2008}:
\begin{equation}
\left| g^{\mathrm{TO}}_{\pi,\mathbf{K};\pi^*,2\mathbf{K}} \right|^2 \propto \lim_{d \to 0} \left( \frac{ \left[ \varepsilon^{\pi^*}_{\mathbf{K}} - \varepsilon^{\pi}_{\mathbf{K}}\right](d) }{ d } \right)^2,
\end{equation}
where $d$ is the amplitude of the displacement and $\varepsilon^{\pi(\pi^*)}_{\mathbf{K}}(d)$ is the energy of the $\pi$ ($\pi^*$) band at $K$ when the atoms are displaced.

When incorporating electron-electron interaction effects on the level of the $GW$~approximation into the band energies, this procedure allows one to calculate the electron-phonon matrix elements at these high-symmetry points while taking into account correlation effects more accurately.
This procedure, however, is limited to high-symmetry points for $\mathbf{k}$ and $\mathbf{q}$ as only there the electron-phonon matrix elements can be related to changes of the band energies.

\subsubsection{Electron-light coupling}

The final ingredient needed for a theoretical description of Raman spectra is the coupling of electrons to light, i.e. an external electromagnetic field.
An external electromagnetic field can be described in terms of a vector potential $\mathbf{A}(\mathbf{r})$ and a scalar potential $\phi(\mathbf{r})$.
In Coulomb gauge $\mathbf{\nabla} \cdot \mathbf{A}(\mathbf{r}) = 0$, the scalar potential must be zero for fields vanishing at infinity.
The external electromagnetic field can therefore be described entirely in terms of a vector potential $\mathbf{A}(\mathbf{r})$.
The Coulomb gauge imposes one constrain on $\mathbf{A}(\mathbf{r})$, leaving two degrees of freedom for the external electromagnetic field, which can be identified with the two possible linearly independent polarization directions of light.

The coupling of electrons to an external vector potential $\mathbf{A}(\mathbf{r})$ can be introduced via the {\it minimal coupling} prescription \footnote{We work with {\it Gaussian} (or {\it cgs}) units. In SI units, the factor $e/c$ is to be replaced by $e$.} $\mathbf{p} \to \mathbf{p} + \frac{e}{c}\mathbf{A}(\mathbf{r})$.
The electronic Hamiltonian then becomes:
\begin{equation}
\begin{split}
H &= \frac{\left[ \mathbf{p} + \frac{e}{c}\mathbf{A}(\mathbf{r}) \right]^2}{2m_e} + V(\mathbf{r};\{\mathbf{R}_0\}) \\
  &= H_0 + \frac{e}{c m_e} \mathbf{A}(\mathbf{r}) \cdot \mathbf{p} + \frac{e^2}{2 c^2 m_e} [\mathbf{A}(\mathbf{r})]^2 \\
  &= H_0 + H^{(1)}_{\mathrm{el-R}} + H^{(2)}_{\mathrm{el-R}}.
\end{split}
\end{equation}
The coupling to the external electromagnetic field generates two more terms in the Hamiltonian that can be treated in perturbation theory.

To obtain the matrix elements for the coupling of electrons to individual photons, the electromagnetic field has to be quantized.
For this, one expands the vector potential in the eigenmodes of a free electromagnetic field, i.e. into plane waves, and promotes it to an operator:
\begin{equation}
\mathbf{A}(\mathbf{r}) = \int \frac{\mathrm{d}^3k}{(2\pi)^3} \sum_{\mu} \sqrt{\frac{4 \pi c^2\hbar}{2\omega_k}}
                                                                  \left[   \boldsymbol{\epsilon}_{\mu,\mathbf{k}} a_{\mu,\mathbf{k}}
                                                                         + \left(\boldsymbol{\epsilon}_{\mu,-\mathbf{k}}\right)^* a^{\dagger}_{\mu,-\mathbf{k}} \right] \mathrm{e}^{i\mathbf{k}\cdot\mathbf{r}},
\end{equation}
where $\mu$ labels the two possible polarizations of a photon, $\omega_k = c |\mathbf{k}|$ and $\boldsymbol{\epsilon}_{\mu,\mathbf{k}}$ are the frequency and polarization vector of a photon with polarization $\mu$ and wave vector $\mathbf{k}$, respectively, and $a^{(\dagger)}_{\mu,\mathbf{k}}$ is an operator that destroys (creates) a photon with polarization $\mu$ and wave vector $\mathbf{k}$.

The matrix elements of the electron-one-photon Hamiltonian between two unperturbed electronic states and with one photon with polarization $\mu$ and wave vector $\mathbf{k}'$ in the initial state then reads:
\begin{equation}
\begin{split}
\gamma^{\mu}_{s,\mathbf{k}-\mathbf{k}';s',\mathbf{k}} &= \langle s,\mathbf{k}-\mathbf{k}'; \mu,\mathbf{k}' | H^{(1)}_{\mathrm{el-R}} | s', \mathbf{k} \rangle \\
                                             &=  \frac{e}{c m_e} \sqrt{\frac{4 \pi c^2\hbar}{2\omega_{k'}}} \boldsymbol{\epsilon}^*_{\mu,\mathbf{k}'} \cdot
                                                 \Big\langle s, \mathbf{k}-\mathbf{k}' \Big| \mathrm{e}^{i\mathbf{k}'\cdot\mathbf{r}}\mathbf{p} \Big| s',\mathbf{k} \Big\rangle,
\end{split}
\end{equation}
where $|s,\mathbf{k}-\mathbf{k}'; \mu,\mathbf{k}'\rangle$ denotes a state consisting of an unperturbed electronic state specified by the band index $s$ and Bloch wave vector $\mathbf{k}-\mathbf{k}'$ and a photon with polarization $\mu$ with wave vector $\mathbf{k'}$.
This expression is often simplified further by the \emph{dipole approximation} $\exp(i \mathbf{k'} \cdot \mathbf{r}) \approx 1$, which is justified by the fact that the wave length of the light ($\sim$300-800~nm for light in the visible spectrum) is much larger than the typical crystal length scale given by the lattice constant ($\sim$2.46~\AA~for graphene), i.e. $\mathbf{k}' \cdot \mathbf{r} \approx 0$ if $\mathbf{r}$ varies on the crystal scale only as it does in the matrix elements between Bloch states.
In other words, for the description of Raman spectra, we can treat the external photons as having zero wave vector (i.e. we set $\mathbf{k}' = 0$), yet finite frequency.
Again, we will collect the matrix elements between the possible combinations of $\pi$ and $\pi^*$~states for fixed $\mathbf{k}$ and light polarization $\mu$ in a 2$\times$2-matrix $\boldsymbol{\gamma}^{\mu}_{\mathbf{k}} = \left(\gamma^{\mu}_{s,\mathbf{k};s',\mathbf{k}}\right)_{s,s'=\pi,\pi^*}$.\\
\\
Finally, one can follow a similar procedure to obtain the matrix elements between $\pi$ and $\pi^*$ states for the electron-two-photon Hamiltonian $H^{(2)}_{\mathrm{el-R}}$.
For a calculation of the Raman spectrum only those matrix elements are important which involve one photon with polarization $\mu$ in the initial state and one photon with polarization $\nu$ in the final state.
We will denote the corresponding 2$\times$2-matrix of matrix elements in the basis of $\pi$ and $\pi^*$ states by $\boldsymbol{\tilde{\gamma}}^{\mu\nu}_{\mathbf{k}}$.

\subsubsection{Electron-defect scattering}

The final building block needed for a complete description of the Raman spectrum within the framework of perturbation theory is a description of electron-defect scattering.
If the graphene sheet contains defects, electrons may scatter from them elastically, leading to additional observable peaks in the Raman spectrum.
Since a first, qualitative understanding of these defect-assisted peaks does not require a detailed model for electron-defect scattering, we will not have a closer look at this issue but merely refer the reader to the relevant literature~\cite{hwang2008b,venezuela2011}.
While we will discuss defect-assisted peaks in the following sections, we will not have a detailed look at its treatment in perturbation theory and hence we will not discuss the modeling of electron-defect scattering here.

\subsubsection{Summary and Feynman rules}

To conclude the section, we summarize the elemental building blocks that are needed for the calculation of the Raman spectrum of graphene.
Each of these elemental processes can be represented graphically by a so-called \emph{Feynman diagram}.
In these diagrams each involved (quasi-)particle is represented by a specific line.
We represent electrons (and holes) by full lines with arrows, phonons by dashed lines, and photons by wavy lines.
The elemental Feynman diagrams for the various building blocks are shown in Table~\ref{tab:Feynman-Rules}.

\newcolumntype{C}{>{\centering\arraybackslash} m{5cm} }
\begin{table}[htb]
	\begin{tabular}{lCl}
		& Diagram & Feynman rule \\
		\hline
		\hline
		\\
		\pbox{5cm}{Propagation\\of electron} &
		\includegraphics{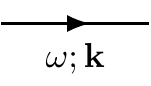} &
		$i\mathbf{G}_{\mathbf{k}}(\omega)$ \\
		\hline
		\\
		\multirow{2}{*}[-1cm]{\pbox{5cm}{Electron-light\\interaction}} &
		\includegraphics{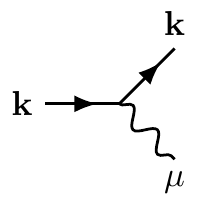} &
		$(-i)\boldsymbol{\gamma}_{\mathbf{k}}^{\nu}$ \\
		&
		\includegraphics{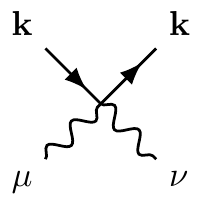} &
		$(-i)\boldsymbol{\tilde{\gamma}}_{\mathbf{k}}^{\mu\nu}$ \\
		\hline
		\\
		\multirow{2}{*}[-1cm]{\pbox{5cm}{Electron-phonon\\interaction}} &
		\includegraphics{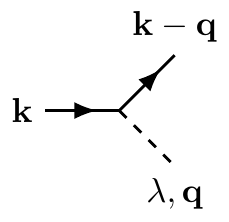} &
		$(-i)\mathbf{g}_{\mathbf{k}-\mathbf{q}}^{\lambda}$ \\
		&
		\includegraphics{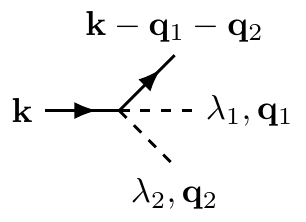} &
		$(-i)\mathbf{\tilde{g}}_{\mathbf{k}-\mathbf{q}_1-\mathbf{q}_2}^{\lambda_1 \lambda_2}$
	\end{tabular}
	\caption{
	{\bf Feynman rules for the basic building blocks for Raman scattering.}
	Feynman diagrams and associated factors appearing in the perturbation series for the Raman scattering amplitudes.
	Each factor is a 2$\times$2-matrix in the basis of $\pi$ and $\pi^*$ states, i.e., each element of the matrix is a matrix element between $\pi$ and $\pi^*$ states.
	}
	\label{tab:Feynman-Rules}
\end{table}

In addition to the matrix elements for scattering processes between particles, there also appears a diagram for the propagation of an unperturbed electron state between two scattering events.
This amplitude is given by the \emph{retarded Green's function}, defined as:
\begin{equation}
G^R_{s,\mathbf{k}}(t) =
\begin{cases}
\mathrm{e}^{-i\varepsilon^s_{\mathbf{k}}t/\hbar} \left( 1 - n_{\mathrm{F}}(\varepsilon^s_{\mathbf{k}}) \right), & t > 0 \\
-\mathrm{e}^{-i\varepsilon^s_{\mathbf{k}}t/\hbar} n_{\mathrm{F}}(\varepsilon^s_{\mathbf{k}}), & t < 0,
\end{cases}
\end{equation}
where $\varepsilon^s_{\mathbf{k}}$ is the energy of the state $|s=\pi,\pi^*,\mathbf{k}\rangle$ and $n_{\mathrm{F}}(\varepsilon)$ denotes the Fermi-Dirac distribution.
The retarded Green's function describes the propagation of an unperturbed electron state for positive times ($t>0$) and of an unperturbed hole state for negative times $t<0$.
Since the perturbation theory calculation will be done in frequency space, we also give the Fourier transform of the retarded Green's function:
\begin{equation}
G_{s,\mathbf{k}}(\omega) =
  \frac{n_\mathrm{F}(\varepsilon^s_{\mathbf{k}})}{\hbar\omega - \varepsilon^s_{\mathbf{k}} - \frac{i}{2}\gamma^s_{\mathbf{k}}}
+ \frac{1 - n_\mathrm{F}(\varepsilon^s_{\mathbf{k}})}{\hbar\omega - \varepsilon^s_{\mathbf{k}} + \frac{i}{2}\gamma^s_{\mathbf{k}}}.
\end{equation}
Here, we accounted for the fact that an electronic state $|s,\mathbf{k}\rangle$ has a finite lifetime $\gamma^s_{\mathbf{k}}$, due to electron-phonon, electron-defect, and electron-electron scattering. As in the sections before, it is convenient to write the Fourier-transformed Green's function as a 2$\times$2-matrix in the space of $\pi$ and $\pi^*$ states at fixed $\mathbf{k}$: $\mathbf{G}_{\mathbf{k}}(\omega) = \left(G_{s,\mathbf{k}}(\omega)\delta_{s,s'}\right)_{s,s'=\pi,\pi^*}$, where $\delta_{s,s'}$ is the Kronecker delta.

\subsection{General kinematic considerations and possible Raman processes}

\begin{figure}[htb]
	\centering
	\includegraphics[scale=0.5]{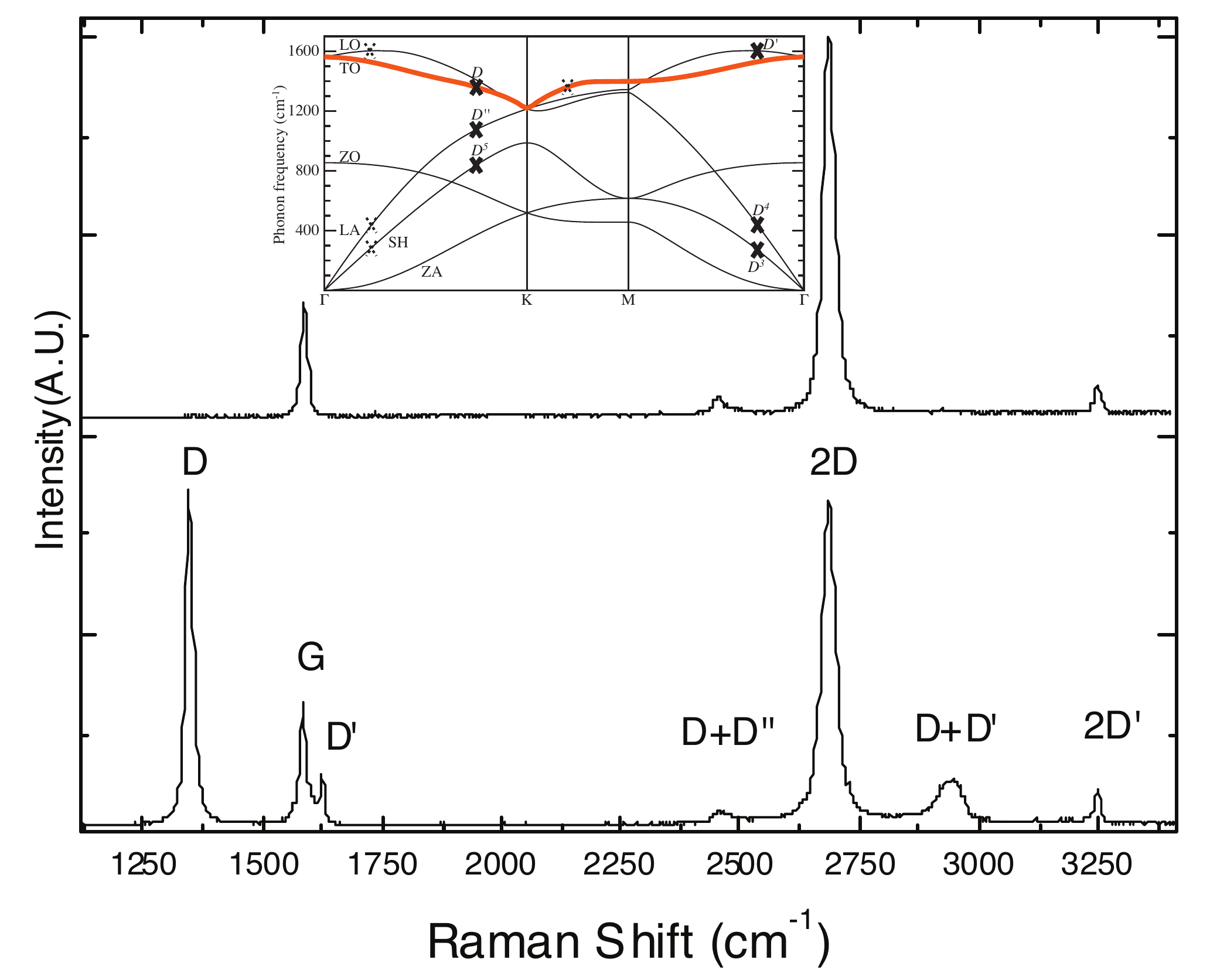}
	\caption{
	{\bf Example of a Raman spectrum and contributing phonons.}
	Two example of a Raman spectrum for a sample without (upper spectrum) and with defects (lower spectrum).
	The most prominent peaks as described in the text have been labeled.
	Inset: Phonon dispersion of graphene, including the corrected TO~branch.
    The phonons which most prominently contribute to the various Raman peaks are marked by crosses.
    (Figure assembled from figures taken from Refs.~\onlinecite{ferrari2013} and \onlinecite{venezuela2011}.)
	}
	\label{fig:example-spectrum}
\end{figure}

Before we use the elementary processes and their associated matrix elements to compute the quantum mechanical amplitude for the various possible Raman processes, we will first discuss some kinematic aspects of Raman processes in general and discuss which Raman processes are expected on kinematic grounds.
In a general Raman process, the initial state consists of one photon of energy $\varepsilon_{\mathrm{in}} = \hbar \omega_{\mathrm{in}}$ and polarization $\mu$ and the final state contains one photon of energy $\varepsilon_{\mathrm{out}} = \hbar \omega_{\mathrm{out}}$ and polarization $\nu$ plus $n$ phonons with wave vectors $\mathbf{q}_1, \ldots, \mathbf{q}_n$ from branches $\lambda_1, \ldots, \lambda_n$.
The energy conservation law determines the observed Raman shift
\begin{equation}
\omega_{\mathrm{in}} - \omega_{\mathrm{out}} = \sum_{i=1}^{n} \omega_{\mathbf{q}_i,\lambda_i}
\end{equation}
as the sum of the frequencies of the involved phonons.
Traditionally, Raman shifts (and phonon frequencies) are converted to shifts of the inverse wavelength, $\Delta \lambda^{-1} = \lambda_{\mathrm{in}}^{-1} - \lambda_{\mathrm{out}}^{-1}$, and given in units of cm$^{-1}$, where the inverse wavelength can be expressed in terms of the frequency $\omega$ via the relation $\lambda^{-1}=\omega/(2 \pi c)$.
For reference, an energy of $\hbar\omega=1$~meV corresponds to a Raman shift of approximately 8~cm$^{-1}$.
The wave vectors of the latter are further restricted by momentum conservation.
In the case that there are no defects in the sample, the sum of the wave vectors of the phonons must be zero, since the wave vectors of photons are negligible and hence don't need to be considered in the momentum conservation law:
\begin{equation}
\sum_{i=1}^{n} \mathbf{q}_i = 0\qquad\text{(no defects)}.
\end{equation}
In the presence of defects, the sum of the momenta of the phonons need not be equal to zero anymore since electronic scattering at defects can make up for it.
If we limit ourselves to a maximum of two phonons, these kinematic considerations lead to the following categorization of Raman processes and their corresponding designations:
\\
\\
\emph{a)} One-phonon processes without defects ($G$ peak)\\
Momentum conservation forces the phonon momentum to be equal to zero, i.e. the phonon must be a phonon from the Brillouin zone center, i.e. the $\Gamma$ point.
Non-zero Raman shifts further require the phonon to be an optical one since the frequencies of acoustic phonons vanish at $\Gamma$.
Furthermore, horizontal mirror symmetry forbids processes with one out-of-plane phonon.
Thus the phonons that contribute to the one-phonon processes in the absence of defects are the two degenerate, optical in-plane phonons at $\Gamma$.
These are doubly degenerate due to rotation symmetry, i.e. they possess the same energy leading to the same Raman shift for both.
The corresponding peak in the Raman spectrum, located at around 1580~cm$^{-1}$ in pristine graphene, is universally called the $G$ peak as it was first observed in graphite.
Note that, as will be discussed in a later section, the breaking of rotation symmetry, for example by strain, leads to the two optical phonons having different frequencies and a splitting of the $G$ peak.
\\
\\
\emph{b)} Two-phonon processes without defects (most prominently the $2D$, $2D'$, and $D+D''$ peaks)\\
Here momentum conservation requires the wave vectors of the involved phonons to be opposite, i.e. $\mathbf{q}_1=\mathbf{q}=-\mathbf{q}_2$.
The vector $\mathbf{q}$, however, is not constrained by any further condition and can in principle correspond to any point in the Brillouin zone.
This explains why the corresponding Raman peaks are much broader than the $G$ peak, as a continuum of phonon frequencies contributes to them.
The fact that there are still clear, observable peaks is related to the fact that certain phonon wave vectors and frequencies lead to a resonance in the Raman amplitudes as will be discussed in the next section.
The peaks that belong to this category are the $2D$ peak, whose dominant contribution comes from two phonons of the TO branch with $\mathbf{q}$ near $K$ as is located at around 2700~cm$^{-1}$, the $2D'$ peak at around 3200~cm$^{-1}$, whose main contribution stems from two phonons of the LO branch near $\Gamma$, and the $D+D''$ peak, which is dominated by one phonon of the TO branch with $\mathbf{q}$ near $K$ and one phonon from the LA branch with opposite wave vector and appears at around 2450~cm$^{-1}$.
\\
\\
\emph{c)} One-phonon processes with defects (most prominently the $D$ and $D'$ peaks)\\
In contrast to the $G$ peak, the presence of defects allows phonons away from the $\Gamma$ point to participate in a one-phonon Raman process as the momentum of the phonon can be compensated by electron-defect scattering (see next section for an example).
Thus, in principle, any phonon can contribute to the Raman spectrum.
However, just as for the peaks in \emph{b)}, the spectrum is dominated by a few selected phonons that lead to a resonance in the Raman amplitude.
The two most prominent peaks that fall into this category are the $D$ peak, which is mostly due to a TO phonon near $K$ and is located at around 1350~cm$^{-1}$, and the $D'$ peak, which is mostly caused by a phonon from the LO branch near $\Gamma$ and occurs at around 1600~cm$^{-1}$.
As their names suggest, these phonons are the same as the ones observed in the two-phonon overtones $2D$ and $2D'$ from the previous paragraph.
\\
\\
\emph{d)} Two-phonon processes with defects (most prominently the $D+D'$ peak)\\
Finally, there are additional two-phonon Raman processes that are made possible by the presence of defects.
In theses processes the two phonons do not need to have opposite wave vectors and can thus stem from different points in the Brillouin zone.
The most prominent peak in this category is the $D+D'$ peak at around 2950~cm$^{-1}$, in which one phonon from the TO branch with a wave vector from near $K$ and an LA phonon with a wave vector close to $\Gamma$ make up the resonant contribution.
\\
\\
A typical, experimentally obtained Raman spectrum for a sample with defects is shown in Fig.~\ref{fig:example-spectrum}, where the most prominent peaks are labeled and the most dominant phonons are identified.

Having given an overview over the kinematic constraints that govern Raman spectrum, we will now turn to the dynamical aspects in the next section and show how to obtain expressions for the Raman amplitudes and how these an be used to determine which phonons play the most important role in the various processes.

\subsection{Diagrammatic calculation of the Raman amplitudes}

In this section we will illustrate the technique of diagrammatic perturbation theory to obtain mathematical expressions for the quantum mechanical amplitudes for the various Raman processes.
The reader can find a more detailed account of this technique in Ref.~\onlinecite{fetter1971,peskin1995,mahan2000}.
We will mainly focus on two examples: One-phonon and two-phonon processes without defects.
Afterward, we will briefly comment on the defect-assisted peaks.

\subsubsection{One-phonon, defect-free processes ($G$~peak)}

\begin{figure}[htb]
	\centering
	\subfloat[]{
		\centering
		\includegraphics{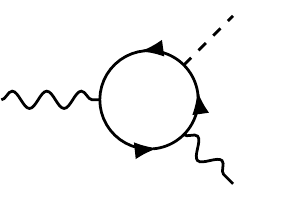}
	}
	\hspace{1cm}
	\subfloat[]{
		\centering
		\includegraphics{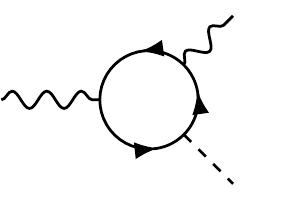}
	}
	\hspace{1cm}
	\subfloat[]{
		\centering
		\includegraphics{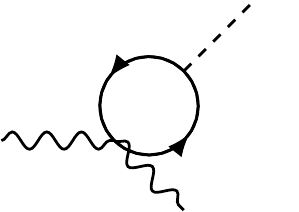}
	}
	\caption{
	{\bf Leading-order Feynman diagrams for the $G$~peak.}
	The three leading-order Feynman diagrams that can be constructed from the basic building blocks of the theory.
	}
	\label{fig:diagrams_Gpeak}
\end{figure}

For the first example, consider a defect-free process with one photon with polarization $\mu$ and energy $\hbar \omega_{\mathrm{in}}$ in the initial state and one photon with polarization $\nu$ and energy $\hbar \omega_{\mathrm{out}}$ plus a phonon of branch $\lambda$ and with momentum $\mathbf{q}$ in the final state.
This is the process that gives rise to the $G$ peak.
The three leading-order, topologically inequivalent Feynman diagrams that can be assembled from the elementary diagrams of Table~\ref{tab:Feynman-Rules} are shown in Fig.~\ref{fig:diagrams_Gpeak}.

Note that momentum conservation dictates $\mathbf{q} = 0$ since the momentum of photons is negligible as discussed earlier.
The wave vectors of the involved virtual electronic states are therefore all the same.
The condition $\mathbf{q}=0$ also fixes the type of the phonon to be optical as acoustic phonons with zero momentum do not couple to $\pi$ and $\pi^*$ states \cite{manes2007}.
The phonon branch index $\lambda$ is thus restricted to the two branches TO and LO which are degenerate at the $\Gamma$ point.
Energy conservation fixes the frequency of the outgoing photon to $\omega_{\mathrm{out}} = \omega_{\mathrm{in}} - \omega_{\mathbf{q}=0,\lambda}$.
The finite lifetime, i.e., decay width, of the phonon can be included via the substitution $\omega_{\mathbf{q}=0,\lambda} \to \omega_{\mathbf{q}=0,\lambda} - i/2 \gamma_{\mathbf{q}=0,\lambda}$, where the relation between lifetime and decay width is given by $\gamma = \hbar/\tau$.
With these kinematic constraints, application of the Feynman rules yields the following mathematical expressions for the amplitudes corresponding to the three diagrams:
\begin{align}
\mathcal{M}^G_1 &= \sum_{\mathrm{spin}} \sum_{\mathbf{k}} \int \frac{\mathrm{d}\omega}{2\pi}\, \mathrm{e}^{+i \omega 0^+}
                   (-1) (-i)^3 i^3 \times
                   \mathrm{tr}\left[ \mathbf{G}_{\mathbf{k}}(\omega) \left(\boldsymbol{\gamma}^{\mu}_{\mathbf{k}}\right)^{\dagger}
                                     \mathbf{G}_{\mathbf{k}}(\omega-\omega_{\mathrm{in}}) \mathbf{g}^{\lambda}_{\mathbf{k},\mathbf{k}}
                                     \mathbf{G}_{\mathbf{k}}(\omega-\omega_{\mathrm{out}}) \boldsymbol{\gamma}^{\nu}_{\mathbf{k}} \right] \\                                 
\mathcal{M}^G_2 &= \sum_{\mathrm{spin}} \sum_{\mathbf{k}} \int \frac{\mathrm{d}\omega}{2\pi}\, \mathrm{e}^{+i \omega 0^+}
                   (-1) (-i)^3 i^3 \times
                   \mathrm{tr}\left[ \mathbf{G}_{\mathbf{k}}(\omega) \boldsymbol{\gamma}^{\nu}_{\mathbf{k}}
                                     \mathbf{G}_{\mathbf{k}}(\omega+\omega_{\mathrm{out}}) \mathbf{g}^{\lambda}_{\mathbf{k},\mathbf{k}}                                       
                                     \mathbf{G}_{\mathbf{k}}(\omega+\omega_{\mathrm{in}}) \left(\boldsymbol{\gamma}^{\mu}_{\mathbf{k}}\right)^{\dagger} \right] \\
\mathcal{M}^G_3 &= \sum_{\mathrm{spin}} \sum_{\mathbf{k}} \int \frac{\mathrm{d}\omega}{2\pi}\, \mathrm{e}^{+i \omega 0^+}
                   (-1) (-i)^2 i^2 \times
                   \mathrm{tr}\left[ \mathbf{G}_{\mathbf{k}}(\omega) \mathbf{g}^{\lambda}_{\mathbf{k},\mathbf{k}}
                                     \mathbf{G}_{\mathbf{k}}(\omega-\omega_{\mathbf{q}=0,\lambda}) \boldsymbol{\tilde{\gamma}}^{\mu \nu}_{\mathbf{k}} \right],
\end{align}
where it is understood that $\omega_{\mathrm{out}}$ is expressed through $\omega_{\mathrm{in}}$ and $\omega_{\mathbf{q}=0,\lambda}$ and the latter is a complex number that includes the finite lifetime of the phonon, as discussed before.
Notice that, in our convention, where the dimensions of the electron-phonon and electron-light matrix elements are those of an energy and the dimension of the Green's function is inverse energy, the dimension of each matrix element is that of a frequency, i.e. a rate.

To obtain the amplitude for each diagram from the elementary Feynman rules, one goes against the fermion flow (i.e. against the direction of the arrows) and writes down, from left to right the factors for each piece one encounters, imposing momentum and energy (frequency) conservation at each vertex.
Since all elementary pieces are 2$\times$2 matrices, the final result will be a matrix.
After adding a factor of $(-1)$ that accompanies a fermion loop, one sums over all internal degrees of freedom, which includes a sum over spin, and takes the trace of the matrix product and also includes summing over any undetermined, internal momenta and frequencies.

The integral over the loop frequency $\omega$ can be evaluated with the residue theorem of complex analysis.
Note that in the leading order diagrams for Raman scattering, only the electronic Green's functions have poles that contribute to the contour integral.
The factor $\exp(i \omega 0^+)$ prescribes that the contour be closed in the upper half of the complex plane and therefore only those electronic states contribute that feature a positive imaginary part, which are those states that are occupied.
Thus the integration over $\omega$ picks out the occupied states.
After the integral has been carried out, the expression for each Feynman diagram becomes a sum of terms.
For example, the amplitude $\mathcal{M}^G_2$ becomes a sum of three terms:
\begin{equation}
\begin{split}
\mathcal{M}^G_2 = \sum_{\mathrm{spin}} \sum_{\mathbf{k}} \left(-\frac{1}{\hbar}\right) \sum_{s=\pi,\pi^*}
                   n_{\mathrm{F}}(\varepsilon^s_{\mathbf{k}}) \Bigg\{
                                      & \left. \left[ \boldsymbol{\gamma}^{\nu}_{\mathbf{k}}
                                      \mathbf{G}_{\mathbf{k}}(\omega+\omega_{\mathrm{out}}) \mathbf{g}^{\lambda}_{\mathbf{k},\mathbf{k}}
                                      \mathbf{G}_{\mathbf{k}}(\omega+\omega_{\mathrm{in}}) \left(\boldsymbol{\gamma}^{\mu}_{\mathbf{k}}\right)^{\dagger} \right]_{ss}
                           \right|_{\hbar \omega = \varepsilon^s_{\mathbf{k}} + \frac{i}{2}\gamma^s_{\mathbf{k}}} \\
                                    + & \left. \left[ \mathbf{g}^{\lambda}_{\mathbf{k},\mathbf{k}}
                                      \mathbf{G}_{\mathbf{k}}(\omega+\omega_{\mathrm{in}}) \left(\boldsymbol{\gamma}^{\mu}_{\mathbf{k}}\right)^{\dagger} 
                                      \mathbf{G}_{\mathbf{k}}(\omega) \boldsymbol{\gamma}^{\nu}_{\mathbf{k}} \right]_{ss}
                           \right|_{\hbar \omega = -\hbar\omega_{\mathrm{out}} + \varepsilon^s_{\mathbf{k}} + \frac{i}{2}\gamma^s_{\mathbf{k}}} \\
                                    + & \left. \left[ \left(\boldsymbol{\gamma}^{\mu}_{\mathbf{k}}\right)^{\dagger}
                                      \mathbf{G}_{\mathbf{k}}(\omega) \boldsymbol{\gamma}^{\nu}_{\mathbf{k}}
                                      \mathbf{G}_{\mathbf{k}}(\omega+\omega_{\mathrm{out}}) \mathbf{g}^{\lambda}_{\mathbf{k},\mathbf{k}} \right]_{ss}
                           \right|_{\hbar \omega = -\hbar\omega_{\mathrm{in}} + \varepsilon^s_{\mathbf{k}} + \frac{i}{2}\gamma^s_{\mathbf{k}}} \Bigg\}
\end{split}
\end{equation}
The number of terms generated in this way is equal to the number of internal propagators (i.e. the lines that make up the loop).
For a simple loop, as it is encountered in the diagrams for the leading order terms for Raman scattering, the number of internal lines is equal to the number of vertices.
This means that a diagram with three external lines that do not meet in a common vertex will lead to a sum of three terms after the loop frequency has been integrated over.
Some books and articles, e.g. the commonly cited Ref.~\onlinecite{yu2010}, prefer to work directly with these terms and write down a diagram for each one of them.
These kinds of diagrams are known as Goldstone diagrams and should not be confused with the Feynman diagrams used in this text.
In fact, every Feynman diagram can be viewed as a short-hand notation for the sum of several Goldstone diagrams with the sum being written as an integral over the loop frequency.
Unfortunately, the distinction between Goldstone and Feynman diagrams is note always made in the literature one sometimes encounters Goldstone diagrams being labeled as Feynman diagrams and {\it vice versa}.
As pointed out, both are essentially equivalent as long as one uses the corresponding set of rules to translate a diagram into its corresponding amplitude and as long as one makes sure to write down all possible topologically non-equivalent diagrams of the chosen approach.
In the Goldstone diagram formulation this entails several different time-ordering of the external lines, while in the Feynman diagram formalism the loop integral takes care of this technicality.
We thus prefer to work with the Feynman diagram formalism as it greatly reduces the number of diagrams that need to be drawn.
In our case, the three Feynman diagrams for the process giving rise to the $G$ peak correspond to eight Goldstone diagrams.\\
\\
Once one has obtained the quantum mechanical amplitudes for the individual diagrams, the total amplitude for the process can be obtained by adding the individual amplitudes:
\begin{equation}
\mathcal{M}_{G\text{ peak}}^{\mu \nu \lambda} (\omega_{\mathrm{in}};\gamma_{\mathrm{el}};\omega_{\mathrm{ph}},\gamma_{\mathrm{ph}}) = \sum_{d=1}^{3}\mathcal{M}_d,
\end{equation}
where we explicitly noted all quantities which the total matrix element depends on.
These are the frequency of the incoming photon, the lifetimes of the involved electronic states (summarily denoted as $\gamma_{\mathrm{el}}$), and the frequency and decay width of the phonon, abbreviated as $\omega_{\mathrm{ph}}$ and $\gamma_{\mathrm{ph}}$, respectively.
Note that the phonon frequency of an optical phonon at $\Gamma$ is independent of the branch $\lambda$, as noted before.
Finally, Fermi's golden rule yields the total rate of the Raman process, i.e., the probability per unit time for the process to happen, which is directly proportional to the recorded Raman intensity:
\begin{equation}
I_G(\omega_{\mathrm{out}})\propto\frac{\mathrm{d}P_{G\text{ peak}}}{\mathrm{d}t} = 2\pi \sum_{\lambda = \text{LO,TO}}\left| \mathcal{M}_{G\text{ peak}}^{\mu \nu \lambda} (\omega_{\mathrm{in}};\gamma_{\mathrm{el}};\omega_{\mathrm{ph}},\gamma_{\mathrm{ph}}) \right|^2 \times \delta(\omega_{\mathrm{in}} - \omega_{\mathrm{out}} - \omega_{\mathrm{ph}})
\end{equation}
For a phonon with finite lifetime, the $\delta$-function has to be replaced by a Lorentzian:
\begin{equation}
\delta(\omega_{\mathrm{in}} - \omega_{\mathrm{out}} - \omega_{\mathrm{ph}})
\to \frac{1}{\pi}\frac{\gamma_{\mathrm{ph}}/2}{(\omega_{\mathrm{in}} - \omega_{\mathrm{out}} - \omega_{\mathrm{ph}})^2 + (\gamma_{\mathrm{ph}}/2)^2}
\end{equation}
From this expression we can immediately see that the position and width of the peak are entirely determined by the frequency (approx. 1580~cm$^{-1}$ in pristine graphene) and width (5-15~cm$^{-1}$ depending on the sample quality) of the phonon.
The intensity of the $G$ peak, however, depends predominantly on the excitation frequency $\omega_{\mathrm{in}}$ as well as on the lifetime of the involved electronic states, while the frequency and lifetime of the phonon only have some minor influence on the intensity of the peak.
Regarding the latter, it should be noted that due to the absence of a band gap in graphene, there are always electronic transitions that are in resonance with the energy of the incoming light.
The condition for resonance is that the energy of the in- or outgoing light matches the energy of an electronic transition.
To give an example of a resonant term appearing in the Raman amplitude, we take a look at one of the terms contained in the matrix product of the first term of $\mathcal{M}^G_2$:
\begin{equation}
\begin{split}
\mathcal{M}^G_2 \supset & \sum_{\mathrm{spin}} \sum_{\mathbf{k}} \left(-\frac{1}{\hbar}\right) n_{\mathrm{F}}(\varepsilon^{\pi}_{\mathbf{k}})
                           \left(1 - n_{\mathrm{F}}(\varepsilon^{\pi^*}_{\mathbf{k}})\right) \left(1 - n_{\mathrm{F}}(\varepsilon^{\pi^*}_{\mathbf{k}})\right) \\
                        & \times \frac{ \langle \pi_{\mathbf{k}}; \omega_{\mathrm{out}}^{\nu}| H^{(1)}_{\mathrm{el-R}} |\pi^*_{\mathbf{k}} \rangle
                                        \langle \pi^*_{\mathbf{k}}; \omega_{\mathrm{ph}}^{\lambda}| H^{(1)}_{\mathrm{el-ph}} |\pi^*_{\mathbf{k}} \rangle
                                        \langle \pi^*_{\mathbf{k}}| H^{(1)}_{\mathrm{el-R}} |\pi_{\mathbf{k}}; \omega_{\mathrm{in}}^{\mu} \rangle }
                                    { \left[ \hbar\omega_{\mathrm{in}} - (\varepsilon^{\pi^*}_{\mathbf{k}}-\varepsilon^{\pi}_{\mathbf{k}}) + \frac{i}{2} (\gamma^{\pi}_{\mathbf{k}}+\gamma^{\pi^*}_{\mathbf{k}}) \right]
                                      \left[ \hbar(\omega_{\mathrm{in}} - \omega_{\mathrm{ph}}) - (\varepsilon^{\pi^*}_{\mathbf{k}}-\varepsilon^{\pi}_{\mathbf{k}}) + \frac{i}{2} (\gamma^{\pi}_{\mathbf{k}}+\gamma^{\pi^*}_{\mathbf{k}} + \gamma_{\mathrm{ph}}) \right] },
\end{split}
\end{equation}
where we replaced the frequency of the outgoing light by the difference of the frequency of the incoming light and the phonon frequency and accounted for the finite lifetime of the phonon by adding a negative imaginary part to the phonon frequency.
This term corresponds to a process where an electron in the $\pi$~band is resonantly excited to the $\pi^*$~band by the incoming photon from which it drops to a virtual state in the $\pi^*$ band by emitting a phonon and then recombines with the hole by emitting the photon found in the final state.
When the resonance condition $\Delta\varepsilon(\mathbf{k}) = \hbar\omega_{\mathrm{in}} \text{ or } \hbar\omega_{\mathrm{out}}$ is met, the denominator becomes minimal, leading to the resonant behavior.
Note that, since the band structure is monotonically increasing away from the Fermi surface (i.e. the $K$ point), the density of states increase monotonically with energy as well and hence the number of states that are in resonance with the energy of the incoming light increases with $\omega_{\mathrm{in}}$, leading to the observed strong behavior of $I_G$ as a function of $\omega_{\mathrm{in}}$~\cite{cancado2007,basko2009}.
By contrast, Raman scattering in semi-conductors with sizable band gaps is mostly non-resonant, unless the excitation energy matches the energy of the band gap or an excitonic state within the gap, whereupon a strong resonance is recorded.\\
\\
In the limit of small $\omega_\mathrm{in}$, the matrix element vanishes, which can easily be explained by the following argument:
For small $\omega_{\mathrm{in}}$, the electronic states that are in resonance with the excitation energy are located very close to the $K$ point.
However, very close to the $K$ point, the band structure of graphene can be considered to be conic (the conic band structure is often called \emph{Dirac cone} due to the Hamiltonian reducing to the Dirac Hamiltonian in this limit).
A conic band structure has a continuous in-plane rotation symmetry, which implies angular momentum conservation.
If we consider circularly polarized \footnote{The case of linearly polarized photons and phonons can be constructed by taking suitable linear combinations of the circular polarized modes.} light and phonons at $\Gamma$, each of them has an angular momentum of $l_z=\pm\hbar$.
The initial state thus has angular momentum $\pm \hbar$, while the final state can have an angular momentum of $+2\hbar$, $0\hbar$, or $-2\hbar$, which would violate angular momentum conservation.
Therefore, in the case of a conic band structure, the $G$ peak process would not be possible in leading order in perturbation theory as it would violate angular momentum conservation.
If one goes one order beyond the linear, conic approximation for the band structure, however, the terms that break the continuous rotation symmetry down to the 120$^{\circ}$ rotation symmetry of the lattice cause a \emph{trigonal warping} of the band structure.
The reduced 120$^{\circ}$ symmetry means that angular momentum is only conserved modulo $\pm3\hbar$, in other words, states with angular momenta that differ by an integer multiple of $3\hbar$ get mixed.
This in turn leads to a non-vanishing Raman amplitude at the leading order in perturbation theory, since now the total angular momenta of the initial and final states can differ by an integer multiple of $3\hbar$ and hence an incoming photon with angular momentum $+\hbar$ can generate a phonon and photon both with angular momentum $-\hbar$.
This argument also illustrates that it is important to go beyond the Dirac cone approximation of the band structure when considering the $G$ peak and that it is necessary to include the trigonal warping terms, which make the $G$ peak process possible in the first place.

\subsubsection{Two-phonon, defect-free processes (e.g., $2D$, $2D'$, and $D+D''$~peaks)}

\begin{figure}[htb]
	\centering
	\subfloat[]{
		\centering
		\includegraphics{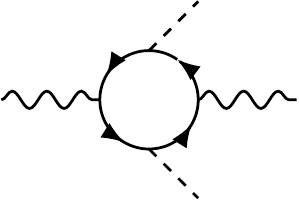}
	}
	\hspace{3cm}
	\subfloat[]{
		\centering
		\includegraphics{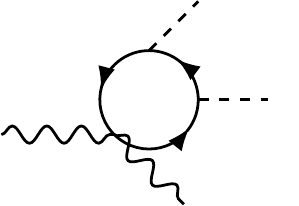}
	}
	\\
	\subfloat[]{
		\centering
		\includegraphics{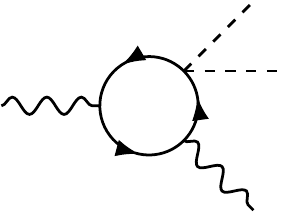}
	}
	\hspace{3cm}
	\subfloat[]{
		\centering
		\includegraphics{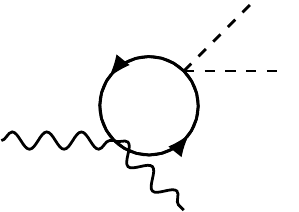}
	}
	\caption{
	{\bf Four examples of leading-order Feynman diagrams for the two-phonon, defect-free peaks.}
	Four of the eleven Feynman diagrams that appear in leading order of perturbation theory.
	Of special importance is the type of diagram shown in a) as it can become \emph{double resonant} (see text).
	}
	\label{fig:diagrams_2Dpeak}
\end{figure}

As a second example for the diagrammatic calculation of the Raman spectrum, consider again a defect-free process, this time with two phonons in the final state.
This kind of process gives rise to the $2D$, $2D'$, and $D+D''$~peaks, among others.
With the available building blocks, the amplitude in lowest order in perturbation theory can be constructed from eleven Feynman diagrams (corresponding to 38 Goldstone diagrams), four arbitrarily chosen ones of which are shown in Fig.~\ref{fig:diagrams_2Dpeak}.

\begin{figure}[htb]
	\centering
	\subfloat[]{
		\centering
		\includegraphics[scale=0.10]{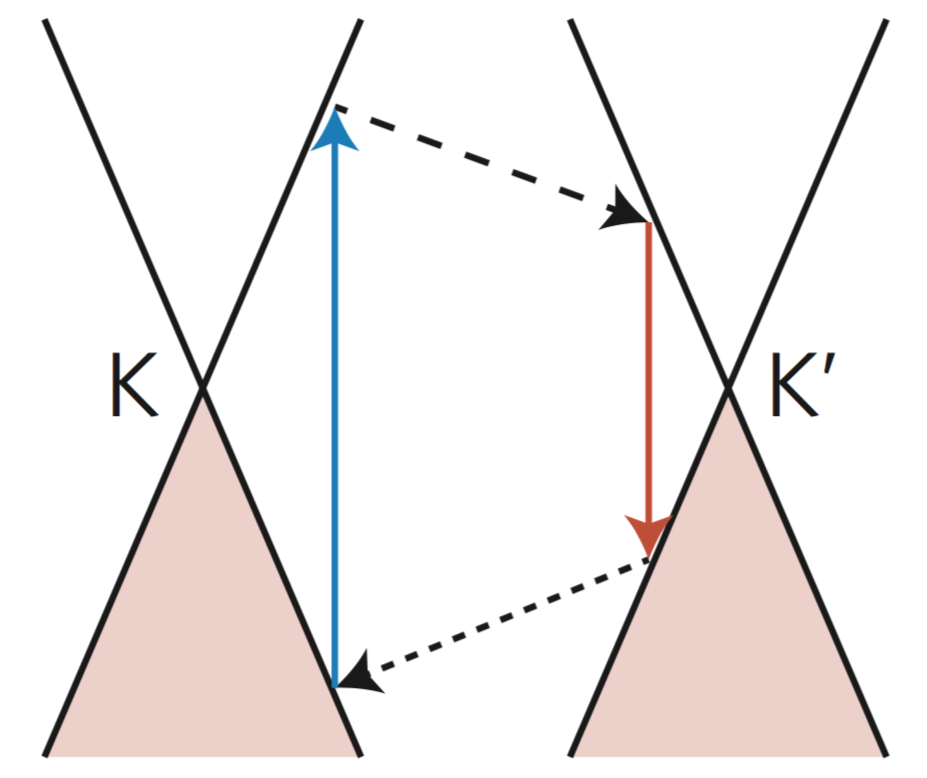}
	}
	\subfloat[]{
		\centering
		\includegraphics[scale=0.10]{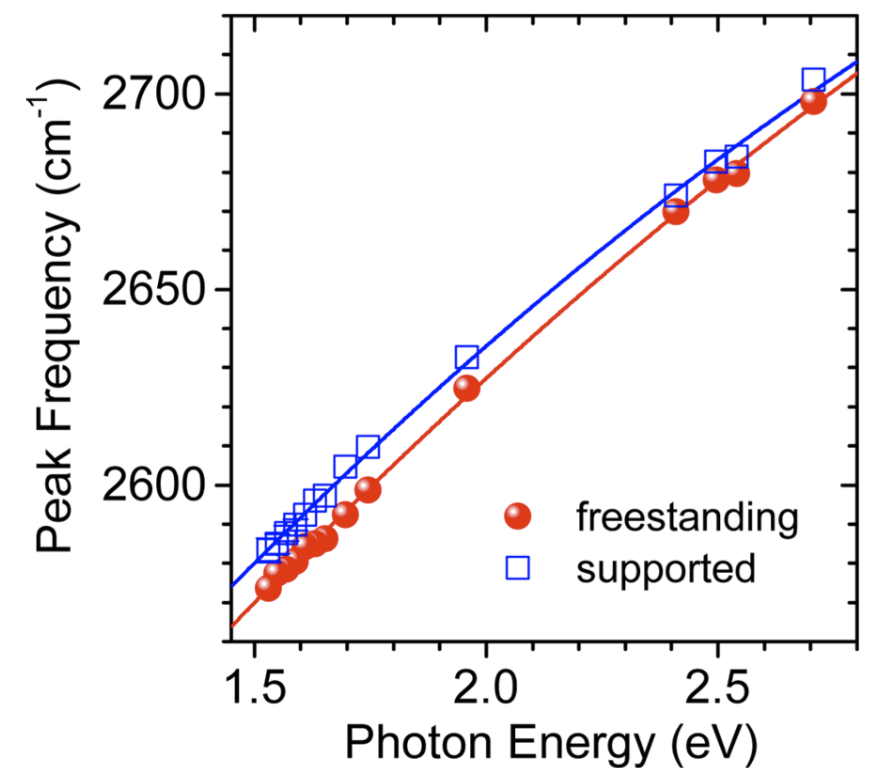}
	}
	\caption{
	{\bf Illustration of one possible double-resonant Raman process and dispersion of the $2D$ line.}
	(a) Illustration of the exemplary microscopic Raman process described in the text, which involves two phonons with wave vectors $\pm\mathbf{q}$ near $K$ and electronic states with wave vectors $\mathbf{k}$ near $K$ and $\mathbf{k}+\mathbf{q}$ near K', respectively.
	Full lines represent the conic electronic band structure around $K$ and $K'$ and the shaded regions represent the filled $\pi$~band.
	The vertical, full arrows represent the excitation and subsequent recombination of an electron-hole pair, while the diagonal, dashed and dotted arrows, represent the scattering of electron and hole, respectively, from a state with wave vector near $K$ to a state with wave vector near $K'$.
	(Figure taken from Ref.~\onlinecite{ferrari2013}.)
	(b) Observed dependence of the position of the $2D$~peak on the energy of the incoming photon.
	Dots and squares refer to data taken on freestanding and supported graphene, respectively.
	Full lines represent fits to the experimental data.
	(Figure taken from Ref.~\onlinecite{berciaud2013}.)
	}
	\label{fig:2D-cones}
\end{figure}

Again, application of the Feynman rules leads to the following expressions for the four example diagrams:
\begin{alignat}{3}
\mathcal{M}^{\text{2 ph.}}_1 &= \sum_{\mathrm{spin}} \sum_{\mathbf{k}} \int \frac{\mathrm{d}\omega}{2\pi}\, \mathrm{e}^{+i \omega 0^+}
                   (-1) (-i)^4 i^4 \times
                   \mathrm{tr}\Bigg[ && \mathbf{G}_{\mathbf{k}}(\omega) \mathbf{g}^{\lambda_1}_{\mathbf{k},\mathbf{k}+\mathbf{q}}
                                        \mathbf{G}_{\mathbf{k}+\mathbf{q}}(\omega+\omega_{\mathbf{q},\lambda_1}) \boldsymbol{\gamma}^{\nu}_{\mathbf{k}+\mathbf{q}} \nonumber \\
              &               && \times \mathbf{G}_{\mathbf{k}+\mathbf{q}}(\omega+\omega_{\mathrm{out}}+\omega_{\mathbf{q},\lambda_1}) \mathbf{g}^{\lambda_2}_{\mathbf{k}+\mathbf{q},\mathbf{k}}
                                        \mathbf{G}_{\mathbf{k}}(\omega+\omega_{\mathrm{in}}) \left(\boldsymbol{\gamma}^{\mu}_{\mathbf{k}}\right)^{\dagger} \Bigg] \\
\mathcal{M}^{\text{2 ph.}}_2 &= \sum_{\mathrm{spin}} \sum_{\mathbf{k}} \int \frac{\mathrm{d}\omega}{2\pi}\, \mathrm{e}^{+i \omega 0^+}
                   (-1) (-i)^3 i^3 \times
                   \mathrm{tr}\Bigg[ && \mathbf{G}_{\mathbf{k}}(\omega) \mathbf{g}^{\lambda_1}_{\mathbf{k},\mathbf{k}+\mathbf{q}}
                                        \mathbf{G}_{\mathbf{k}+\mathbf{q}}(\omega+\omega_{\mathbf{q},\lambda_1}) \mathbf{g}^{\lambda_2}_{\mathbf{k}+\mathbf{q},\mathbf{k}} \nonumber \\
              &               && \times \mathbf{G}_{\mathbf{k}+\mathbf{q}}(\omega+\omega_{\mathbf{q},\lambda_1}+\omega_{\mathbf{q},\lambda_2}) \boldsymbol{\tilde{\gamma}}^{\mu \nu}_{\mathbf{k}} \Bigg] \\
\mathcal{M}^{\text{2 ph.}}_3 &= \sum_{\mathrm{spin}} \sum_{\mathbf{k}} \int \frac{\mathrm{d}\omega}{2\pi}\, \mathrm{e}^{+i \omega 0^+}
                   (-1) (-i)^3 i^3 \times
                   \mathrm{tr}\Bigg[ && \mathbf{G}_{\mathbf{k}}(\omega) \mathbf{\tilde{g}}^{\lambda_1 \lambda_2}_{\mathbf{k},\mathbf{k}}
                                        \mathbf{G}_{\mathbf{k}}(\omega+\omega_{\mathbf{q},\lambda_1}+\omega_{\mathbf{q},\lambda_2}) \boldsymbol{\gamma}^{\nu}_{\mathbf{k}} \nonumber \\
              &               && \times \mathbf{G}_{\mathbf{k}}(\omega+\omega_{\mathrm{in}}) \left(\boldsymbol{\gamma}^{\mu}_{\mathbf{k}}\right)^{\dagger} \Bigg] \\
\mathcal{M}^{\text{2 ph.}}_4 &= \sum_{\mathrm{spin}} \sum_{\mathbf{k}} \int \frac{\mathrm{d}\omega}{2\pi}\, \mathrm{e}^{+i \omega 0^+}
                   (-1) (-i)^2 i^2 \times
                   \mathrm{tr}\Bigg[ && \mathbf{G}_{\mathbf{k}}(\omega) \mathbf{\tilde{g}}^{\lambda_1 \lambda_2}_{\mathbf{k},\mathbf{k}}
                                        \mathbf{G}_{\mathbf{k}}(\omega+\omega_{\mathbf{q},\lambda_1}+\omega_{\mathbf{q},\lambda_2}) \boldsymbol{\tilde{\gamma}}^{\mu \nu}_{\mathbf{k}} \Bigg],
\end{alignat}
The total amplitude for this process can again be constructed by summing all (in the leading order eleven) amplitudes:
\begin{equation}
\mathcal{M}_{\text{2 ph.}}^{\mu \nu \lambda_1 \lambda_2} (\mathbf{q};\omega_{\mathrm{in}};\gamma_{\mathrm{el}};\omega_{\mathrm{ph}},\gamma_{\mathrm{ph}}) = \sum_{d=1}^{11}\mathcal{M}_d,
\end{equation}
The Intensity is then calculated form Fermi's golden rule after summing over all possible phonon branches $\lambda_1, \lambda_2$ and wave vectors $\mathbf{q}$:
\begin{equation}
I_{\text{2 ph.}}(\omega_{\mathrm{out}})\propto\frac{\mathrm{d}P_{\text{2 ph.}}}{\mathrm{d}t} = 2\pi \sum_{\lambda_1,\lambda_2} \sum_{\mathbf{q}} \left| \mathcal{M}_{\text{2 ph.}}^{\mu \nu \lambda_1 \lambda_2} (\mathbf{q};\omega_{\mathrm{in}};\gamma_{\mathrm{el}};\omega_{\mathrm{ph}},\gamma_{\mathrm{ph}}) \right|^2 \times \delta(\omega_{\mathrm{in}} - \omega_{\mathrm{out}} - \omega_{\mathbf{q},\lambda_1} - \omega_{\mathbf{q},\lambda_2}).
\end{equation}
Again, for phonons with finite lifetimes, the $\delta$-function is to be replaced by a Lorentzian.
From the above expression, we can see that the two-phonon, defect-free process will result in a Raman spectrum that consists of the sum of Lorentzians, one for each combination of phonon branches $\lambda_1, \lambda_2$ and for each phonon wave vector $\mathbf{q}$.
The position of each of these Lorentzians in the Raman spectrum (i.e., the recorded Raman shift) is given by the sum of the frequencies of the two involved phonons.
Since the phonon frequencies for the different wave vectors $\mathbf{q}$ form a continuum, one does not record individual peaks, however, but rather a continuous distribution in the Raman spectrum.
Note that the weight of each Lorentzian, i.e. the weight of the contribution of each wave vector $\mathbf{q}$ and of each of the branches $\lambda_1$ and $\lambda_2$, is given by the corresponding Raman matrix element, which depends again, on the excitation frequency $\omega_{\mathrm{in}}$, the electronic lifetimes $\gamma_{\mathrm{el}}$ and the phonon frequencies and lifetimes $\omega_{\mathrm{ph}}$ and $\gamma_{\mathrm{ph}}$, respectively.\\
\\
As will be discussed in detail in the next paragraph, for each pair $(\lambda_1,\lambda_2)$ of phonon branches considered, the Raman amplitude will have a dominating maximum due to a double resonance for specific phonon wave vectors $\mathbf{q}^{\lambda_1,\lambda_2}_{\mathrm{res}}(\omega_{\mathrm{in}})$ that depend on the excitation frequency.
These will be the phonons that contribute the most to the Raman amplitude and hence the corresponding Lorerentzians will have a much bigger weight than those of the other phonons.
This in turn leads to the observation of a clear peak for the two-phonon processes.
This peak will have its maximum at the center of the Lorentzian corresponding to the $\mathbf{q}^{\lambda_1,\lambda_2}_{\mathrm{res}}$, i.e. at $\omega_{\mathbf{q}_{\mathrm{res}},\lambda_1}+\omega_{\mathbf{q}_{\mathrm{res}},\lambda_2}$.
While the Raman matrix element for these specific wave vectors might be maximal, this does not mean, however, that other ``nearby'' phonons do not sizably contribute to the observed Raman amplitude.
In fact, the condition that fixes $\mathbf{q}^{\lambda_1,\lambda_2}_{\mathrm{res}}$ depends on the excitation energy, which in case of a resonance in the Raman amplitude is equal to some specific electronic excitation energy.
Since the electronic levels are smeared out by a finite decay width $\gamma_{\mathrm{el}}$, so is the resonance condition for $\mathbf{q}^{\lambda_1,\lambda_2}_{\mathrm{res}}$.
Thus, the range of the phonon wave vectors that give a sizable contribution to the Raman spectrum is determined by the electronic lifetime $\gamma_{\mathrm{el}}$ and since the range of $\mathbf{q}$ determines the range of phonon frequencies that contribute to the two-phonon peaks, the electronic lifetime has a major impact on the width of two-phonon peaks.
Note that $\gamma_{\mathrm{el}} = \mathcal{O}(100 \text{ meV}) \gg \gamma_{\mathrm{ph}} = \mathcal{O}(15 \text{ meV})$ and hence the phononic lifetimes hardly play any role for the width of the two-phonon peaks.
This should be contrasted with the width of the $G$ peak which is entirely determined by the widths of the degenerate phonon involved and thus is much narrower than the two-phonon-induced peaks.
For example, typical widths of the $2D$~peak range from 20-40~cm$^{-1}$ compared to the typical widths of 5-15~cm$^{-1}$ for the $G$~peak.~\cite{neumann2015b}\\
\\
To understand the mentioned double-resonant behavior of the Raman amplitude for the two-phonon processes, we take a look at one of the amplitudes corresponding to one of the Feynman diagrams.
In particular, we will look at one of the terms generated by the contour integration over $\omega$ that are contained in the matrix product in $\mathcal{M}^{\text{2 ph.}}_1$:
\begin{equation}
\begin{split}
\mathcal{M}^{\text{2 ph.}}_1 \supset  & \sum_{\mathrm{spin}} \sum_{\mathbf{k}} \left(-\frac{1}{\hbar}\right)n_{\mathrm{F}}(\varepsilon^{\pi}_{\mathbf{k}})
                                        n_{\mathrm{F}}(\varepsilon^{\pi}_{\mathbf{k}+\mathbf{q}}) \left(1 - n_{\mathrm{F}}(\varepsilon^{\pi^*}_{\mathbf{k}})\right)
                                        \left(1 - n_{\mathrm{F}}(\varepsilon^{\pi^*}_{\mathbf{k}+\mathbf{q}})\right) \\
                &      \times  \frac{ \langle \pi_{\mathbf{k}+\mathbf{q}}; \omega_{\mathrm{out}}^{\nu}| H^{(1)}_{\mathrm{el-R}} |\pi^*_{\mathbf{k}+\mathbf{q}} \rangle
                                      \langle \pi_{\mathbf{k}}; \omega_{\mathbf{q}}^{\lambda_1}| H^{(1)}_{\mathrm{el-ph}} |\pi_{\mathbf{k}+\mathbf{q}} \rangle
                                      \langle \pi^*_{\mathbf{k}+\mathbf{q}}; \omega_{-\mathbf{q}}^{\lambda_2}| H^{(1)}_{\mathrm{el-ph}} |\pi^*_{\mathbf{k}} \rangle
                                      \langle \pi^*_{\mathbf{k}}| H^{(1)}_{\mathrm{el-R}} |\pi_{\mathbf{k}}; \omega_{\mathrm{in}}^{\mu} \rangle }
                                    { \left[ \hbar\omega_1 - (\varepsilon^{\pi}_{\mathbf{k}+\mathbf{q}}-\varepsilon^{\pi}_{\mathbf{k}}) + \frac{i}{2} (-\gamma^{\pi}_{\mathbf{k}+\mathbf{q}}+\gamma^{\pi}_{\mathbf{k}}-\gamma_1) \right]
                                      \left[ \hbar(\omega_{\mathrm{out}} + \omega_1) - (\varepsilon^{\pi^*}_{\mathbf{k}+\mathbf{q}}-\varepsilon^{\pi}_{\mathbf{k}}) + \frac{i}{2} (\gamma^{\pi}_{\mathbf{k}}+\gamma^{\pi^*}_{\mathbf{k}+\mathbf{q}}-\gamma_1) \right] } \\
                & \times \frac{1}{ \left[ \hbar\omega_{\mathrm{in}} - (\varepsilon^{\pi^*}_{\mathbf{k}}-\varepsilon^{\pi}_{\mathbf{k}}) + \frac{i}{2} (\gamma^{\pi}_{\mathbf{k}}+\gamma^{\pi^*}_{\mathbf{k}}) \right] }.
\end{split}
\label{eq:2D-amp-term}
\end{equation}
It can be interpreted as describing the excitation of an electron with wave vector $\mathbf{q}$ from the $\pi$ to the $\pi^*$ band due to the absorption of a photon with frequency $\omega_{\mathrm{in}}$ and polarization $\mu$, followed by scattering of the excited electron to a state with wave vector $\mathbf{k}+\mathbf{q}$ in the $\pi^*$ band by emitting a phonon with frequency $\omega_2 \equiv \omega_{-\mathbf{q},\lambda_2}$.
In the meantime the empty state at wave vector $\mathbf{k}$ in the $\pi$ band has been filled by an electron that was scattered from a state with wave vector $\mathbf{k}+\mathbf{q}$ 
in the $\pi$ band by emitting a phonon with frequency $\omega_1 \equiv \omega_{\mathbf{q},\lambda_1}$, leaving behind a hole.
Finally, this hole recombines with the scattered excited electron in the $\pi^*$ band by via emission of a photon with frequency $\omega_{\mathrm{out}}$ and polarization $\nu$.
This process is illustrated in Fig.~\ref{fig:2D-cones}a for two phonons of the same branch (TO) with wave vectors $\pm\mathbf{q}$ near $K$ and electronic states with wave vectors $\mathbf{k}$ near $K$ and $\mathbf{k}+\mathbf{q}$ near $K'$, respectively.

The amplitude for this process becomes maximal when one or several of the denominators become minimal, i.e. when their real part vanishes.
While the amplitude for the $G$~peak could only become singly-resonant since both denominators could not vanish simultaneously, here the presence of electronic states with different wave vectors allows for two of the denominators to vanish simultaneously if the phonon wave vector obeys some resonance condition.
Note that irrespective of the phonon wave vector $\mathbf{q}$, single resonance is always possible, as already discussed for the $G$ peak.
To derive the condition for double resonance for the phonon wave vector $\mathbf{q}$, we look at the amplitude for fixed $\mathbf{q}$ and fixed phonon branches $\lambda_1$ and $\lambda_2$ and require two of the three factors in the denominator of Equation~\ref{eq:2D-amp-term} to vanish for some electronic wave vector $\mathbf{k}$.
We will see that this is only possible for a specific choice of $\mathbf{q}$, which we already denoted as $\mathbf{q}^{\lambda_1,\lambda_2}_{\mathrm{res}}$ in the previous paragraph.
In principle, any two combinations of the three denominators can vanish at once.
We thus restrict ourselves to one of the three possible pairs of vanishing terms to illustrate how to find the resonance condition for the phonon wave vector $\mathbf{q}$.
The two terms we seek to minimize are the first and third of the factors in the denominator.
By setting the real part of the latter to zero, we immediately find a condition for the wave vector of the electronic states that are in resonance with the light:
\begin{equation}
\hbar\omega_{\mathrm{in}} = \varepsilon^{\pi^*}_{\mathbf{k}} - \varepsilon^{\pi}_{\mathbf{k}} \quad \Rightarrow \quad \mathbf{k} = \mathbf{k}_{\mathrm{res}}(\omega_{\mathrm{in}}).
\end{equation}
Note that since $\mathbf{k}$ is a two-dimensional vector, this condition describes an entire resonance surface (a closed line in two dimensions) of $k$-vectors in the Brillouin zone.
For example, in the limit of a conic band structure, the set of $\mathbf{k}_{\mathrm{res}}$ describes a circle in the Brillouin zone.
By also requiring that the first factor in the denominator is minimal, we finally arrive at a condition for $\mathbf{q}$:
\begin{equation}
\hbar\omega_{\mathbf{q},\lambda_1} = \varepsilon^{\pi}_{\mathbf{k}_{\mathrm{res}}(\omega_{\mathrm{in}})+\mathbf{q}} - \varepsilon^{\pi}_{\mathbf{k}_{\mathrm{res}}(\omega_{\mathrm{in}})}
\quad \Rightarrow \quad \mathbf{q} = \mathbf{q}_{\mathrm{res}}(\omega_{\mathrm{in}}).
\end{equation}
Again, $\mathbf{q}$ is a two-dimensional vector and thus this equations defines a resonance surface for $\mathbf{q}_{\mathrm{q}}$, which, for example, is a circle for conic band structure and a conic phonon dispersion as found for $\mathbf{q}$ near $K$ or $\Gamma$.
Due to the shape of the graphene bands and the shape of the optical phonon dispersion near $K$, solutions for these equations always exist.
By plugging the resonant wave vectors $\mathbf{k}_{\mathrm{res}}$ and $\mathbf{q}_{\mathrm{res}}$ and the constraints into the remaining factor that we did not explicitly minimize, we find by using $\omega_{\mathrm{in}} = \omega_{\mathrm{out}} + \omega_1 + \omega_2$:
\begin{equation}
\begin{split}
  & \hbar(\omega_{\mathrm{out}} + \omega_{\mathbf{q}_{\mathrm{res}},\lambda_1}) - ( \varepsilon^{\pi^*}_{\mathbf{k}_{\mathrm{res}}+\mathbf{q}_{\mathrm{res}}} - \varepsilon^{\pi}_{\mathbf{k}_{\mathrm{res}}} ) \\
= & -\hbar\omega_{\mathbf{q}_{\mathrm{res}},\lambda_2} + \varepsilon^{\pi^*}_{\mathbf{k}_{\mathrm{res}}} - \varepsilon^{\pi^*}_{\mathbf{k}_{\mathrm{res}}+\mathbf{q}_{\mathrm{res}}}.
\end{split}
\end{equation}
In case the two phonons stem from them same branch, i.e., $\lambda_1 = \lambda_2$, this equation can be further simplified by plugging in the constraint for $\omega_{\mathbf{q},\lambda_1}$:
\begin{equation}
\begin{split}
  & -\hbar\omega_{\mathbf{q}_{\mathrm{res}},\lambda_2} + \varepsilon^{\pi^*}_{\mathbf{k}_{\mathrm{res}}} - \varepsilon^{\pi^*}_{\mathbf{k}_{\mathrm{res}}+\mathbf{q}_{\mathrm{res}}} \\
= & (\varepsilon^{\pi}_{\mathbf{k}_{\mathrm{res}}} + \varepsilon^{\pi^*}_{\mathbf{k}_{\mathrm{res}}}) - (\varepsilon^{\pi}_{\mathbf{k}_{\mathrm{res}}+\mathbf{q}_{\mathrm{res}}} + \varepsilon^{\pi^*}_{\mathbf{k}_{\mathrm{res}}+\mathbf{q}_{\mathrm{res}}}).
\end{split}
\end{equation}
Notice that the terms in parenthesis are a measure for the electron-hole asymmetry at $\mathbf{k}_{\mathrm{res}}$ and $\mathbf{k}_{\mathrm{res}}+\mathbf{q}_{\mathrm{res}}$.
Since $\mathbf{k}_{\mathrm{res}}$ and $\mathbf{k}_{\mathrm{res}}+\mathbf{q}_{\mathrm{res}}$ have to be near $K$ for the resonance conditions to be fulfilled~\cite{maultzsch2004,venezuela2011}, the third factor will be small as well since the electron-hole asymmetry is small near $K$.

On the other hand, for two phonons of different branches, i.e. $\lambda_1 \neq \lambda_2$, the third factor will in general not be small and this explains why the Raman peaks due to phonons of the same branch are much more visible than the Raman bands with two phonons of different branches.

It should be noted that one can construct similar conditions for $\mathbf{q}^{\lambda_1,\lambda_2}_{\mathrm{res}}$ by demanding that the two other possible pairs of terms be minimal.
In general, this will result, however, in slightly different wave vectors $\mathbf{q}^{\lambda_1,\lambda_2}_{\mathrm{res}}$.
Also note that the resonant conditions resulting from the other diagrams have $\mathbf{q}$ depend on $\lambda_2$ or even on both $\lambda_1$ and $\lambda_2$, which justifies the notation $\mathbf{q}^{\lambda_1,\lambda_2}_{\mathrm{res}}$.
By explicitly evaluating the double-resonance conditions, one finds~\cite{venezuela2011} that the phonons that give the biggest contributions to the various Raman peaks are a phonon from the TO branch with $\mathbf{q}$ near $K$ for the $D$~peak, and phonon from the LO branch with $\mathbf{q}$ near $\Gamma$ for the $D'$~peak and a phonon from the LA branch with $\mathbf{q}$ near $K$ for the $D''$~peak.\\
\\
Finally, we want to point out that $\mathbf{q}^{\lambda_1,\lambda_2}_{\mathrm{res}}$ depends on the excitation frequency $\omega_{\mathrm{in}}$ as mentioned before.
Since the phonon wave vector determines the resonant contribution and hence the position of the maximum of the peak, the two-phonon peaks will shift when the excitation frequency $\omega_{\mathrm{in}}$ is varied, which leads to the these peaks being called \emph{dispersive}.
In the approximation of a linear band structure (i.e. a Dirac cone) and a linear phonon dispersion near $K$ and $\Gamma$ and for the resonant electron and phonon wave vectors being parallel along high-symmetry lines, the two conditions for double resonance become linear equations for $|\mathbf{k}|$ and $|\mathbf{q}|$, which are also linear in $\omega_{\mathrm{in}}$.
Hence the solutions to these equations $|\mathbf{k}_{\mathrm{res}}|$ and $|\mathbf{q}_{\mathrm{res}}|$ are also linear in $\omega_{\mathrm{in}}$ and because of the approximately  linear dispersion of the phonons near $K$ and $\Gamma$, the resonant phonon frequency and with it the peak position are linear in $\omega_{\mathrm{in}}$ as well.
This linear dispersion of the two-phonon frequencies has been observed experimentally, e.g., in Refs.~\onlinecite{thomsen2000,maultzsch2004,berciaud2013} and is shown in Fig.~\ref{fig:2D-cones}b.
Note that contrary to the process for the $G$~peak, the Raman amplitude does not vanish for a conic band structure, since the final state contains three particles (one photon, two phonons), so that angular momentum conservation as implied by a conic band structure can always be obeyed.\\
\\
So far we have focused on the double-resonant contribution to the Raman amplitude and determined which phonon wave vectors $\mathbf{q}$ maximize it, which in turn determines the Raman shift at which the two-phonon Raman peaks are observed.
However, we want to stress that while the double-resonant terms are undoubtedly of high importance for a correct description of the Raman peaks, quantum interference effects also play a crucial role in the correct description of the Raman spectrum~\cite{venezuela2011}.
While the contributions of different phonons are added after squaring the matrix element (different phonons mean different final states), the sum over electronic wave vectors $\mathbf{k}$ has to be done before squaring the matrix element.
As pointed out in the literature~\cite{venezuela2011}, the doubly-resonant electronic states $\mathbf{k}_{\mathrm{res}}$ can constructively or destructively interfere with one another, which can significantly influence the peak shape and height.

\subsubsection{Defect-assisted peaks (e.g., the $D$, $D'$, and $D+D'$~peaks)}

After the detailed discussion of defect-free processes in the last two subsections, we conclude this section by briefly addressing the defect-assisted processes.
In the presence of defects, electron-phonon scattering events can be complemented or replaced by elastic electron-defect scattering events.
For example, if one replaces one of the two electron-phonon scattering events in the defect-free two-phonon process by an electron-defect scattering event, one obtains a process that has one instead of two phonons in the final state.
This phonon can in principle have an arbitrary wave vector $\mathbf{q}$, as already discussed in the section on kinematic constraints.
A simple way to obtain the amplitude for such a process is to replace the electron-phonon(-light) matrix element that involves the second phonon by one of the defect-scattering amplitudes and set the corresponding phonon frequency in the denominator to zero.

While the double resonance conditions that involve the second phonon can thus no longer be fulfilled, the double resonance conditions that involve the first phonon, however, are unaffected by this.
Therefore, the corresponding Raman process will still be doubly resonant for the same phonon wave vectors $\mathbf{q}_{\mathrm{res}}$, we found for the two-phonon, defect-free case above.
Accordingly, the maximum of the corresponding Raman peak will be located at the position of the frequency of the resonant wave vectors.
Compared to the defect-free two-phonon process involving two phonons of the same branch, where the Raman shift of the peak maximum is given by the sum of the frequencies of the two phonons, the Raman shift for the corresponding one-phonon, defect-assisted peak will thus be halved.
This explains the designation $2D$ and $2D'$ for two of the two-(same-)phonon peaks as their corresponding defect-assisted peaks are denoted $D$ (for resonant phonons of the TO branch with $\mathbf{q}$ near $K$) and $D'$ (for resonant phonons of the LA branch near $\Gamma$), respectively.
Since the double resonance conditions are the same for the two-phonon process and one-phonon process with defects, the position of the Raman peaks of the latter also depend on the excitation frequency $\omega_{\mathrm{in}}$.

In addition to the matrix element squared, the intensity of the defect-assisted peaks will also contain a factor of the defect concentration~\cite{venezuela2011}.
This direct linear dependence of the height of the peak on the defect concentration is easily understandable since the height of the peak is proportional to the probability per unit time of the corresponding Raman process.
Since the latter requires a defect to be non-zero, the probability per unit time for the process to happen is proportional to the defect concentration and hence the peak height is as well.
A more detailed discussion of the dependence of the intensity of theses peaks can be found in the final section of this chapter.

Finally, there are also Raman processes which involve two phonons and additional electron-defect scattering events.
This allows for more and different conditions for double resonance, leading to several Raman sidebands and peaks, the most prominent of which is the $D+D'$ peak.

\section{Influence of internal and external parameters on the Raman spectrum}
\label{sec:applications}

In the final section of this chapter, we discuss how various internal properties (e.g. defects or layer number) of a graphene sample or externally controllable parameters (e.g. doping or strain) influence the Raman spectrum.
We will restrict ourselves to the discussion of the arguably most common and arguably best understood external influences, namely the number of layers, the defect concentration, doping, and strain.
Other influences not covered here include disorder and edge effects~\cite{ferrari2007}, substrate influences~\cite{berciaud2013,forster2013,neumann2015b}, and temperature~\cite{calizo2007}, for which the interested reader is referred to the literature.

\subsection{Layer number}

In the previous parts of this chapter, we were only concerned with monolayer graphene.
While it is still by far the most often studied graphene system in contemporary research and allows one to illustrate the general theoretical approach to the calculation of Raman spectra, multi-layer graphene also has received some attention in the literature.
Indeed, one important application of Raman spectroscopy is the determination of the number of layers during sample exfoliation or growth.
We therefore briefly summarize the most important changes to the Raman spectrum if the sample consists not of one but of several layers of graphene.

By increasing the number of layers, the number of atoms in the unit cell changes by 2 for each additional layer.
The number of atoms in the unit cell determines the number of (optical) phonon branches.
For $N$ atoms in the unit cell the number of optical branches is given by $3N-3$, two thirds of which are in-plane phonons.
However, almost across the whole Brillouin zone, the new phonon branches that are relevant for Raman scattering are degenerate with the ones already existing in monolayer graphene.
Only near the $K$ point do the elsewhere degenerate TO branches split slightly~\cite{herziger2014}.

By contrast, the electronic band structure, undergoes a more severe change.
Both the $\pi$ and $\pi^*$ bands split into $N$ bands each.
As the different layers of multilayer graphene are bound mostly by van der Waals~forces, which are much weaker than the intralayer covalent forces, the splitting of the additional bands is small compared to the energy scale of the bands.
Thus all of these bands now play a role in the Raman processes.

By taking both the changes to the phonon dispersion and the electronic band structure into account, we can summarize the changes to the various Raman processes and peaks as follows:\\
\\
For the one-phonon, defect-free process, i.e. the $G$~peak, very little change 
is observed as the position and decay width of the degenerate phonon at $\Gamma$ are as good as unchanged.
The separation in energy of the additional bands from the monolayer ones is so large that the additional possible electronic transitions have transition energies out of range of the phonon.
Hence they couple only weakly to it, which means that the frequency and decay width of the phonon at $\Gamma$ are not measurably different from the monolayer ones.

\begin{figure}[htb]
	\centering
	\subfloat[]{
		\centering
		\includegraphics[scale=0.15]{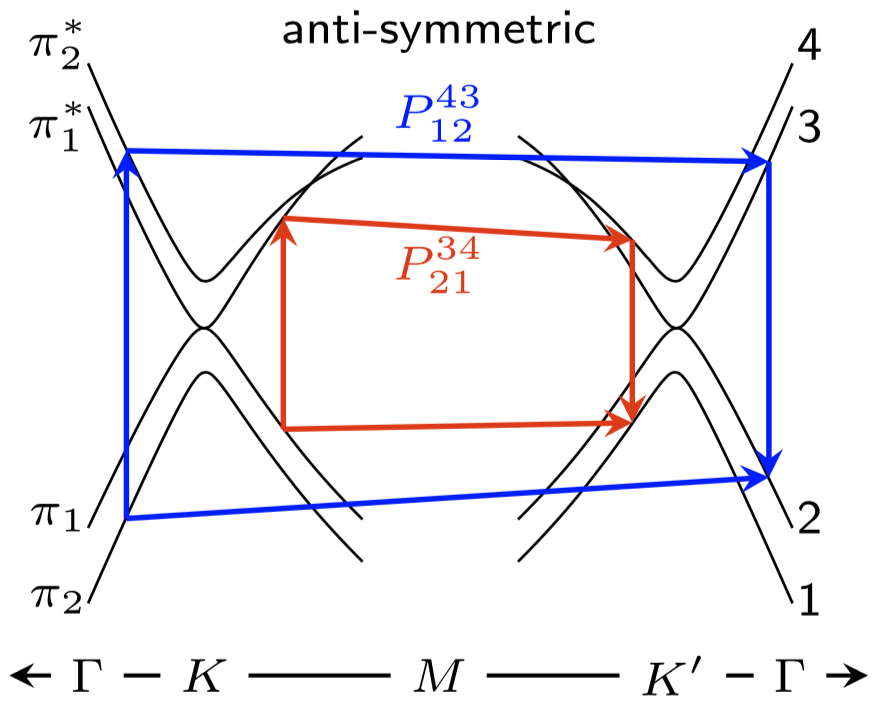}
	}
	\subfloat[]{
		\centering
		\includegraphics[scale=0.15]{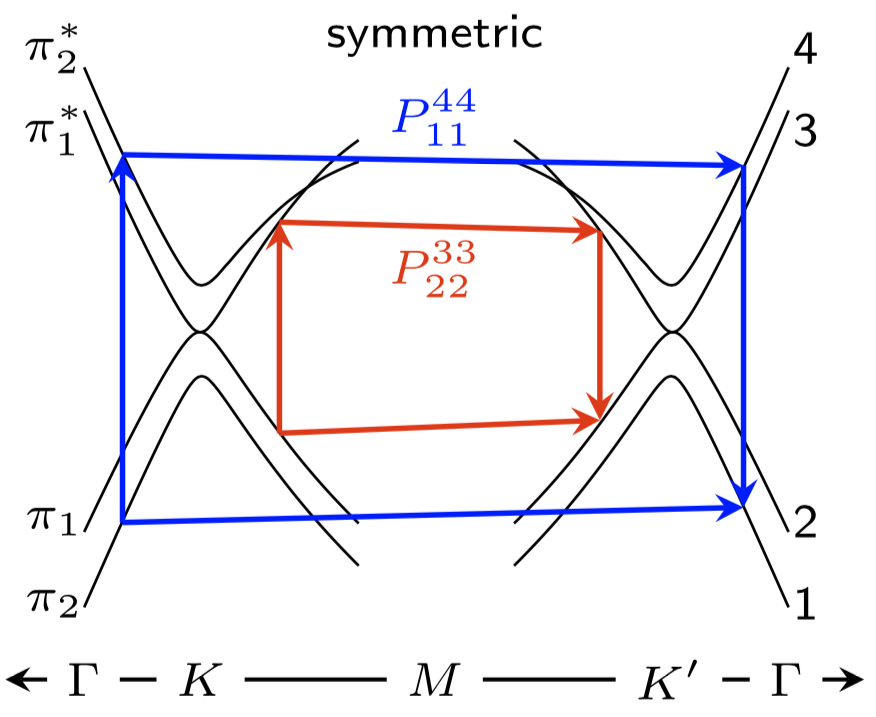}
	}
	\subfloat[]{
		\centering
		\includegraphics[scale=0.2]{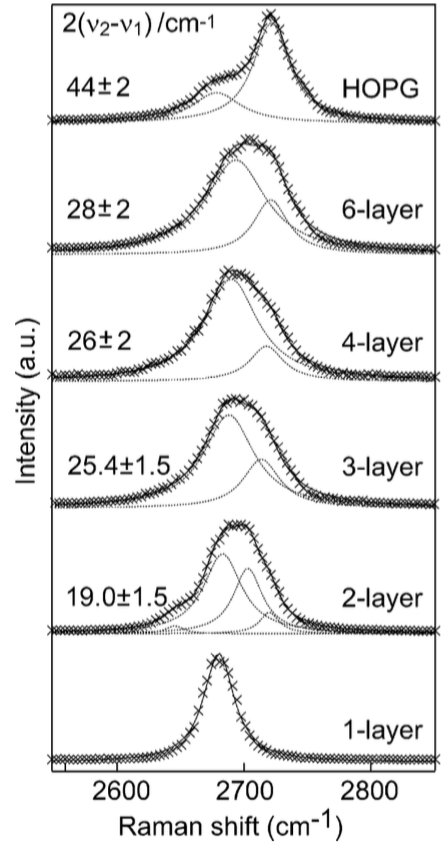}
	}
	\caption{
	{\bf Change of the $2D$~peak width with layer number.}
	(a,b) Two examples of double-resonant processes for bilayer graphene.
	    Shown are the two $\pi$ and $\pi^*$~bands of bilayer graphene and examples of electronic transitions that can go double-resonant.
	    (Figure taken from Ref.~\onlinecite{herziger2014}.)
	(c) Evolution of the FWHM of the $2D$~peak with the number of layers, from monolayer graphene to highly oriented pyrolithic graphite (HOPG).
	    (Figure taken from Ref.~\onlinecite{graf2007}.)
	}
	\label{fig:layer-number}
\end{figure}

For the two-phonon and/or defect-assisted processes, however, the situation is different.
Recall that the existence of a clear peak in monolayer graphene is attributed to the fact that for certain electron wave vectors $\mathbf{k}_{\mathrm{res}}$ and phonon wave vectors $\mathbf{q}_{\mathrm{res}}$ the Raman amplitude becomes double resonant.
The additional electronic bands in multilayer graphene allow for additional double-resonance conditions with slightly different solutions $\mathbf{k}_{\mathrm{res}}$ and $\mathbf{q}_{\mathrm{res}}$ (compare Fig.~\ref{fig:layer-number}a for an example of a double-resonant process in the case of bilayer graphene).
In addition, the slight splitting of the TO branch of the phonon dispersion for $\mathbf{q}$ near $K$ further allows for different $\mathbf{q}_{\mathrm{res}}$.
Altogether, this leads to more phonon wave vectors $\mathbf{q}$ that satisfy a double resonance condition and hence several close-by phonon frequencies now contribute significantly to the observed Raman shift.
This leads to a measurable broadening of the two-phonon and defect-assisted peaks, in particular to a splitting/broadening of the most prominent peak, the $2D$~peak.
As expected from the above considerations, the broadening of these peaks increases with every additional number of layers~\cite{ferrari2006,graf2007} (see Fig.~\ref{fig:layer-number}b) and converges towards the peak structure of bulk graphite.

The broadening of the $2D$~peak in particular is nowadays routinely used to reliably estimate the number of layers during sample exfoliation or growth.

\subsection{Defect concentration}

As already mentioned before, the presence of defects in a graphene sample changes the Raman spectrum in two ways:
On the one hand, it leads to more peaks in the spectrum as the defect-assisted Raman processes become possible.
On the other hand, it modifies the height and widths of the non-defect-assisted peaks.
For both groups of Raman peaks, we will briefly discuss the dependence of their height and width on the defect concentration in this subsection.\\
\\
For the \emph{defect-assisted} peaks, e.g. the $D$, $D'$, and $D+D'$~peaks, the peak height and width depends non-monotonically on the defect concentration $n_D$.

Firstly, there is the direct dependence of the intensity on $n_D$, which appears as a scaling factor in the intensity. This factor represents the fact that the probability for a defect-assisted Raman process is proportional to the number of defects.

Secondly, there is the indirect dependence of the quantum mechanical amplitude on $n_D$ through the decay widths, i.e., the lifetimes, of the intermediate electronic states, $\gamma_{\mathrm{el}} = \gamma_{\mathrm{el}}(n_D)$, which appear in the denominator of the matrix element of the Raman process and thus higher widths lead to smaller quantum mechanical amplitudes.
In general, the decay widths of electronic states contains several contributions related to the different ways in which an electron can scatter.
The three most prominent ones are electron-electron interaction, i.e., Coulomb scattering with another electron, electron-phonon interaction, i.e., activation of a lattice vibration, and scattering from a defect.
The latter contribution to the total decay width of an electronic state will be proportional to $n_D$, as more defects leads to a higher probability for an electron to scatter from a defect.

Depending on the defect concentration, the first or the second aspect will dominate~\cite{venezuela2011}.

For high defect concentrations, the electronic decay widths will be dominated by electron-defect scattering and hence will be proportional to $n_D$.
Since the electronic decay widths appear in each of the multiple factors in the denominator of the Raman matrix element, this effect will eventually outweigh the overall linear factor of $n_D$ appearing in the expression for the Raman intensity.
Therefore, for large enough densities, the heights of the defect-assisted peaks will decrease.

While the peak heights thus behave non-monotonically as a function of $n_D$, the peak widths monotonically increase with $n_D$.
Here, the only relevant effect is the increase of the electronic decay widths with $n_D$.
As discussed before, the broadening of the electronic states leads to a smearing of the resonance condition and hence a wider range of phonon frequencies will non-negligibly  contribute to the Raman peaks.
Thus, the widths of the peaks increase monotonically with defect concentration.\\
\\
The properties of the \emph{non-defect-assisted} peaks on the other hand, depend on the defect concentration in a purely monotonic fashion.

\begin{figure}[htb]
	\centering
	\includegraphics[scale=0.15]{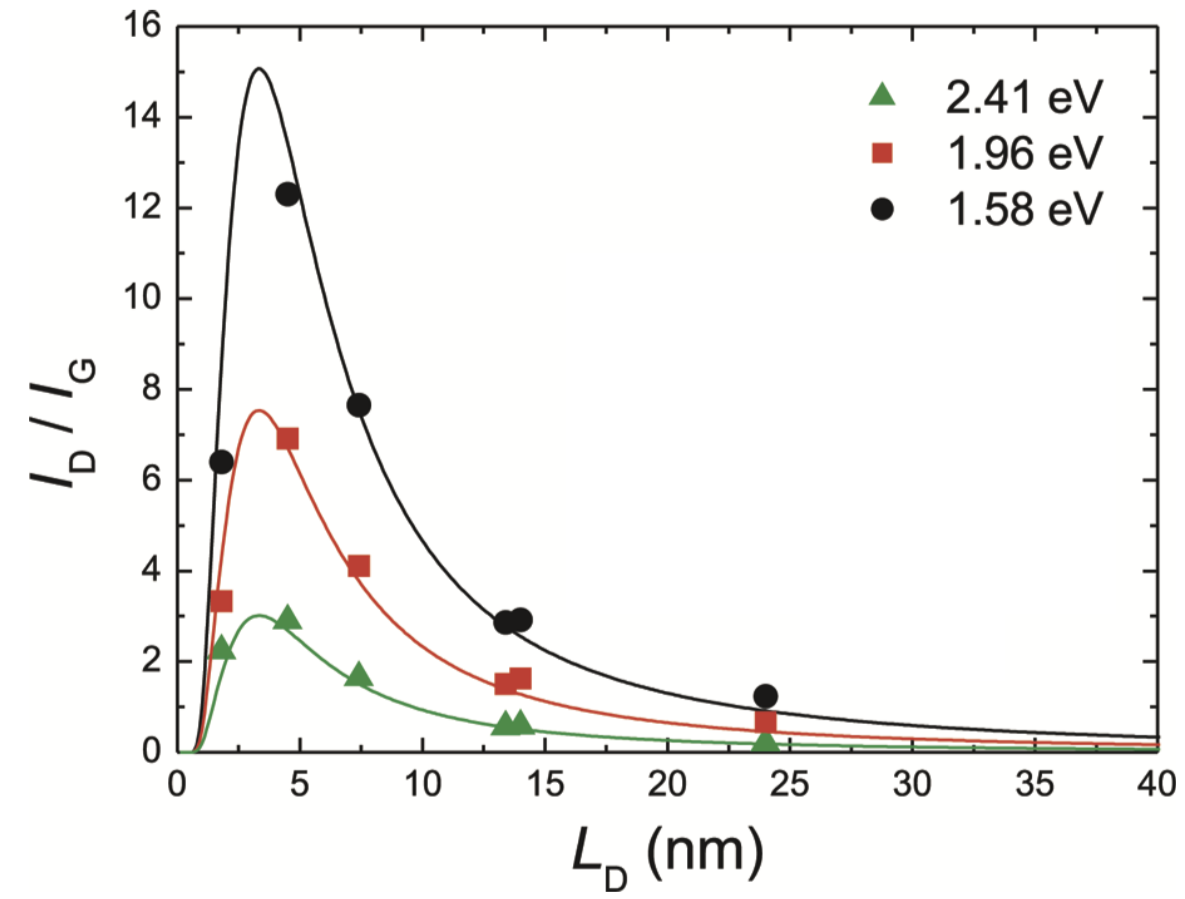}
	\caption{
	{\bf Dependence of the intensity of the Raman peaks on the defect concentration.}
	Ratio of the intensity of the $D$ and $G$ peaks as a function of the average distance between defects $L_D$, which is a measure for the defect concentration $n_D$.
	(Figure taken from Ref.~\onlinecite{cancado2011}.)
	}
	\label{fig:defect-influences}
\end{figure}

Their intensities monotonically decrease as a function of the defect concentration, since they only depend on $n_D$ through the decay widths of the electronic states, which enter the quantum mechanical matrix element through factors in the denominator.
Since for low defect concentrations, the dependence of the electronic decay widths on $n_D$ is negligible, the height of the non-defect-assisted peaks is independent of the defect concentration as well for low $n_D$.
The different behavior of the intensity of the defect-assisted and non-defect-assisted peaks as a function of $n_D$ has been used to estimate the amount of defects and disorder in a sample by measuring the ratio of the intensity of the $D$ and $G$~peaks~\cite{cancado2011} (see Fig.~\ref{fig:defect-influences}).

The behavior of the widths of the defect-free peaks under changes of $n_D$ depends on the specific peak.
The width of the $G$~peak, as shown in the previous section, does not depend on the electronic decay widths as it is purely determined by the lifetime of the created phonon.
The widths of the multi-phonon peaks, such as the $2D$, $2D'$, or $D+D''$~peaks, by contrast, increases with $n_D$, for the same reason as for the defect-assisted peaks.

\subsection{Doping}

Doping, i.e., changes of the charge carrier density, leads to a shift of the Fermi level.
Hole ($p$-type) doping leads to states in the $\pi$~band not being occupied, while electron ($n$-type) doping causes states in the $\pi^*$~band to be occupied.
This will influence the Raman spectrum in three ways:\\

\begin{figure}[htb]
	\centering
	\subfloat[]{
		\centering
		\includegraphics[scale=0.2]{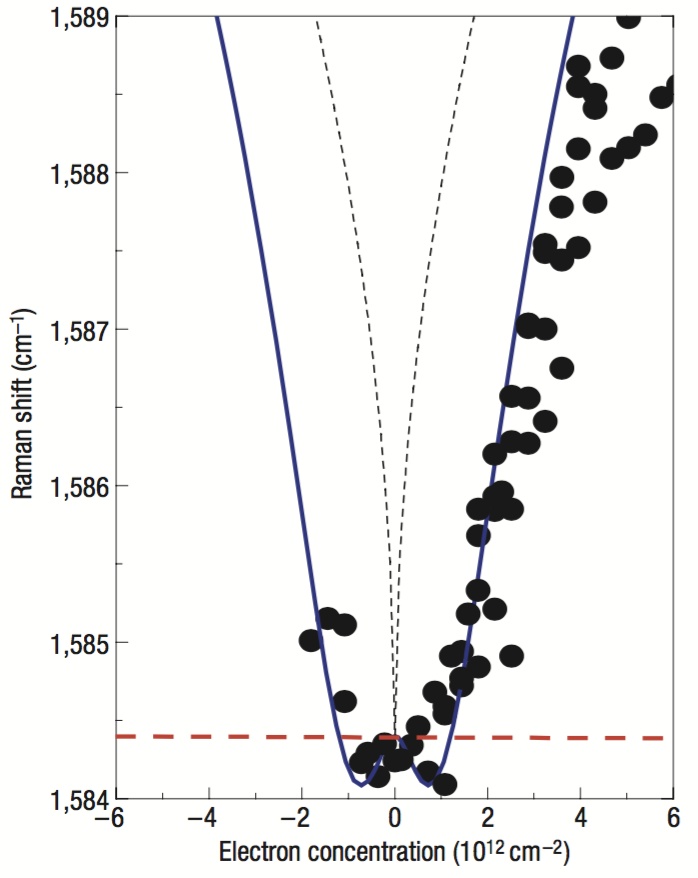}
	}
	\subfloat[]{
		\centering
		\includegraphics[scale=0.2]{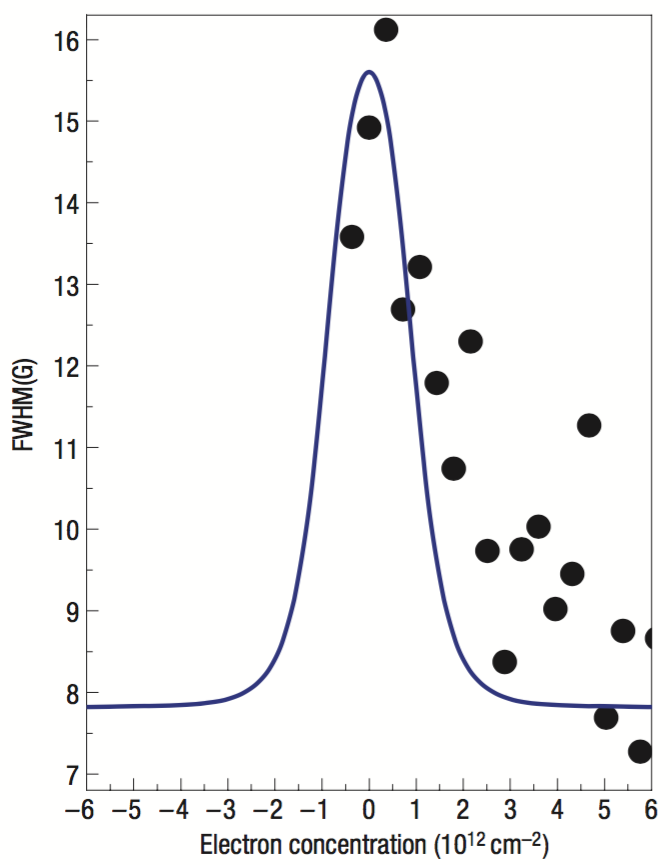}
	}
	\subfloat[]{
		\centering
		\includegraphics[scale=0.2]{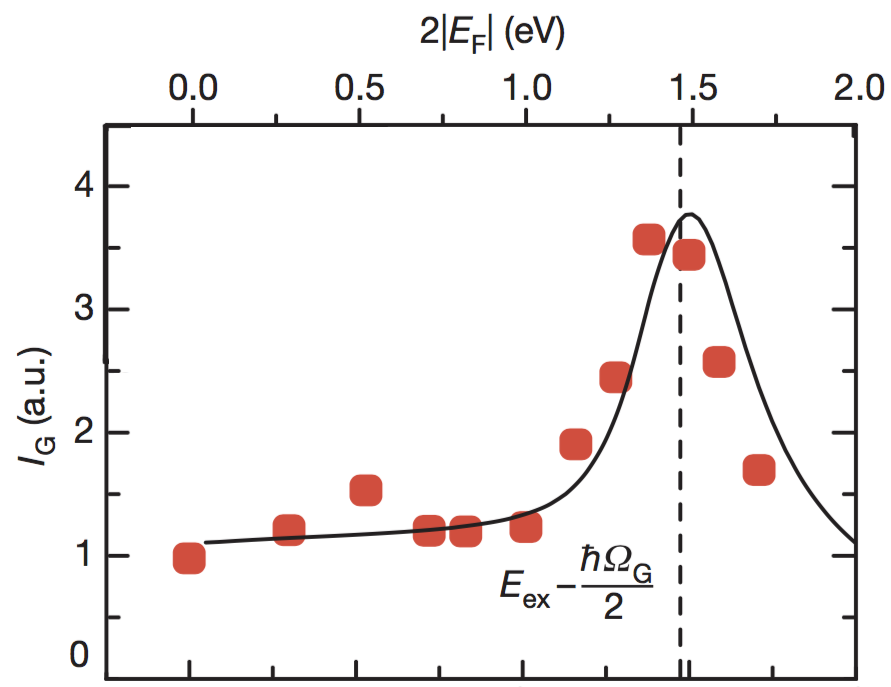}
	}
	\caption{
	{\bf Effects of doping on electrons, phonons, and Raman processes.}
	(a,b) Change of the frequency (a) and the decay width (b) of the in-plane optical phonons at $\Gamma$ with doping, which determine the frequency (position) and width of the Raman $G$~peak.
	     The dots represent experimental data, while the full lines represent the results of a theoretical calculation.
	    (Figures taken from Ref.~\onlinecite{pisana2007}.)
	(c) Intensity of the Raman $G$~peak as a function of Fermi energy, i.e., doping.
	    The dots represent experimental data, while the full line represent the result of a theoretical calculation using a simplified version of the theory presented in the previous section.
  	    (Figure taken from Ref.~\onlinecite{chen2011}.)
	}
	\label{fig:doping-influences}
\end{figure}

Firstly, the electronic band structure is slightly affected as the screening of the electron-electron interaction is increased by the presence of additional charge carriers.
The most important effect of this is decrease of the slope of the $\pi$ and $\pi^*$ bands near the $K$ point~\cite{elias2011}.
This effects the two-phonon and/or defect-assisted Raman processes by changing the condition for double resonance.

An additional effect is the increase of the decay width of excited electronic states (or equivalently a decrease of their lifetime) $\gamma_{\mathrm{el}}$, as the scattering rate of an electron due to electron-electron interaction increases with additional charge carriers in the system.
This will affect the height of the Raman peaks and correspondingly their area, since they are proportional to the matrix element squared of the corresponding Raman process, which in turn decreases for higher decay widths of the electronic states as they appear in the denominator of the amplitude.
Since the amplitude for the two-phonon processes (e.g., the $2D$~peak process) contains one more electronic Green's functions than the amplitude for the one-phonon, defect-free process (leading to the $G$~peak), it contains one more factor of $\gamma_{\mathrm{el}}$, and hence the height and area of the two-phonon peaks reduces much more with doping than the height and area of the $G$~peak~\cite{basko2009b}.
In addition to the height, the width of the two-phonon and/or defect-assisted peaks changes as well as experimentally observed for the $2D$~peak~\cite{berciaud2013}, since it is also related to the electronic decay width, as mentioned in the last section.\\
\\
Secondly, the phonon frequencies and widths change.
This is most noticeable for those phonons which are affected by Kohn anomalies: the degenerate optical in-plane phonons at $\Gamma$ and the TO branch at and around $K$.
The phonon at $\Gamma$ is affected by doping through a change of its \emph{self-energy}, i.e., its change of frequency due to electron-phonon interaction.
The commonly associated picture behind this is the phonon's temporary decay into an electron-hole pair which recombines to produce the same phonon again.
This process is only relevant if the electron or the hole have not been scattered in the meantime, i.e., if they are long-lived enough or, equivalently, if their decay width is low enough. Otherwise this virtual process does not play a role as recombination to the original phonon is not possible.
The decay width is 'low enough' if it is much lower than the transition energy of the electron-hole pair which is equal to the energy of the phonon.
To put it the other way around, for this process to play a role, the lifetime of the excited electron or the hole must be much longer than than the period of the phonon oscillation.
If this is the case, the nuclei are oscillating while the electron stays in its exciting state without 'instantly' dropping back to the ground state and without adiabatically following the nuclei, as assumed in the adiabatic Born-Oppenheimer approximation.
In graphene, the decay widths of the exited electron and the hole are indeed much lower compared to the phonon energy, leading to the statement that the adiabatic Born-Oppenheimer approximation breaks down in graphene~\cite{pisana2007}.
With doping, the occupation of the excited states and holes is changed and this in turn changes the self-energy of the phonon, resulting in an increase of the phonon frequencies of the degenerate optical branches at $\Gamma$~\cite{ando2006,pisana2007}, which means that the position of the $G$~peak shifts towards higher frequencies, as shown in Fig.~\ref{fig:doping-influences}a.
Note that in other materials, the ratio of electronic decay widths and phonon frequencies is usually much larger than in graphene and hence the exited electrons and holes have time to relax to the ground state and adiabatically follow the oscillating nuclei and hence the self-energy correction in other materials is negligible.

At the same time, doping also blocks a possible decay channel of the phonons, by preventing the decay into an electron-hole pair due to the Pauli principle.
If the Fermi level exceeds half of the phonon energy, than the decay into an electron and a hole with each carrying away half of the phonon energy becomes impossible due to the Pauli exclusion principle since the electronic state that would be occupied after the decay is already occupied due to doping.
Thus, if the Fermi level exceeds half of the phonon energy, the phonon decay widths drops noticeably (compare Fig.~\ref{fig:doping-influences}b), as the decay channel into electron-hole pairs makes up a sizable part of the total decay width of the phonons.
In the case of finite temperature, this sharp drop of the decay width is smeared out with the Fermi-Dirac distribution.
The strong decrease of the decay width of the phonon at $\Gamma$ with doping directly amounts to a strong decrease of the width of the Raman $G$~peak, since the latter is equal to the decay width of the phonon, as explained in the theoretical section.
By monitoring the $G$~peak width, one thus has a qualitative measure to check if the Fermi level exceeds half of the phonon energy.\\
\\
For the two-phonon and/or defect-assisted peaks, the physical picture is roughly the same as for the $G$~peak.
As mentioned earlier, the phonons involved in the processes leading to the $D$, $D''$, $D+D''$ and, most importantly, the $2D$~peak have wave vectors near $K$.
In the vicinity of $K$, effects of the Kohn anomaly play an important role since it leads to steepening of the slope of the phonon dispersion, which plays an important role in determining which phonons obey the double-resonance conditions and hence where the center of the Raman peaks is situated.
The Kohn anomaly is very sensitive to the Fermi surface and thus a change of the Fermi level through doping means that it has a large influence on the phonon dispersion around $K$ and consequently on the position of the Raman peaks.
Indeed, it is found experimentally that the two-phonon and/or defect-assisted peaks~\cite{das2008} shift with doping, albeit not as strongly as the $G$~peak.

The decay width of the phonons that fulfill the double-resonance condition, on the other hand, are not affected for low doping levels.
The finite momentum of these phonons does not allow the phonon to decay into an electron-hole pair due to energy and momentum conservation, as the necessary hole states would have to have an energy of around half of the laser energy for which the phonon fulfills the double-resonance condition.
For typical experimental laser energies of 2-3~eV, it would require a Fermi level shift of 1-1.5~eV to block the decay width of the phonons involved in the doubly resonant process.
The influence of doping on the decay widths of the phonons can thus be neglected for typical doping levels achieved in experiment.\\
\\
Finally, doping influences the Raman amplitudes directly.
As seen in the previous section, the Raman amplitudes contain factors of the Fermi-Dirac distribution of the involved electronic states.
By changing the Fermi level and thus varying the occupation of the electronic states, some of the Raman amplitudes are set to zero.
For the $G$~peak amplitude, which does not rely primarily on the resonant terms to exist, doping in general has only minor effects.
It has been demonstrated experimentally however~\cite{chen2011}, that if the Fermi level comes close to half of the frequency of the outgoing light, the total amplitude for the $G$~peak process increases dramatically.
The underlying reason for this increase is the blocking of terms in the amplitude that destructively interfere with the resonant terms, leading to a large \emph{increase} of the amplitude, as depicted in Fig.~\ref{fig:doping-influences}c.

For the two-phonon and/or defect-assisted processes, the situation is different.
Here, the interference between the different terms in the amplitude is believed to be constructive~\cite{chen2011} and therefore raising the Fermi level to half of the frequency of the light only leads to a sharp \emph{decrease} of the peak height and ultimately to its disappearance, as has been demonstrated for the $2D$~peak.\\
\\
We can summarize the effects of doping on the Raman spectrum as follows:
The $G$~peak is expected to shift to a higher frequency.
In addition, it will narrow once the Fermi level is bigger than $\hbar\omega_{\mathrm{ph}}/2$ or lower than $-\hbar\omega_{\mathrm{ph}}/2$, with the width of the peak decreasing to as low as 4-5~cm$^{-1}$~\cite{neumann2015b}.
The two-phonon and/or defect-assisted peaks, in particular the $2D$~peak, are expected to shift as well, albeit not as strong as the $G$~peak.
Furthermore the ratio of the areas of the $2D$ and $G$~peaks decreases as well with doping.

\subsection{Strain}

Mechanical strain has a large influence on the Raman spectrum.
This is hardly surprising as it changes the bond lengths of the atoms, which leads to both a change in the phonon frequencies as the effective 'springs' between the atoms are softened or hardened and to a change in the electronic band structure since the unit cell becomes distorted and hence the Brillouin zone does as well.

\begin{figure}[htb]
	\centering
	\subfloat[]{
		\centering
		\includegraphics[scale=0.15]{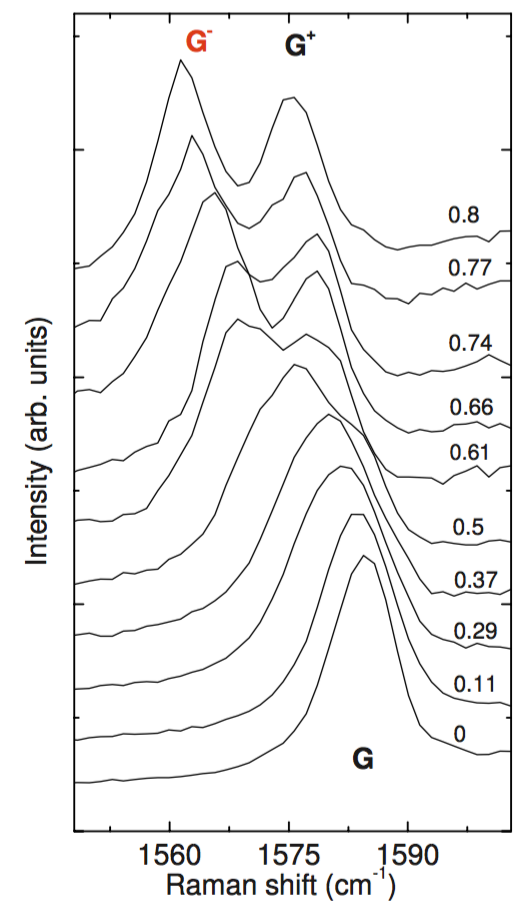}
	}
	\subfloat[]{
		\centering
		\includegraphics[scale=0.15]{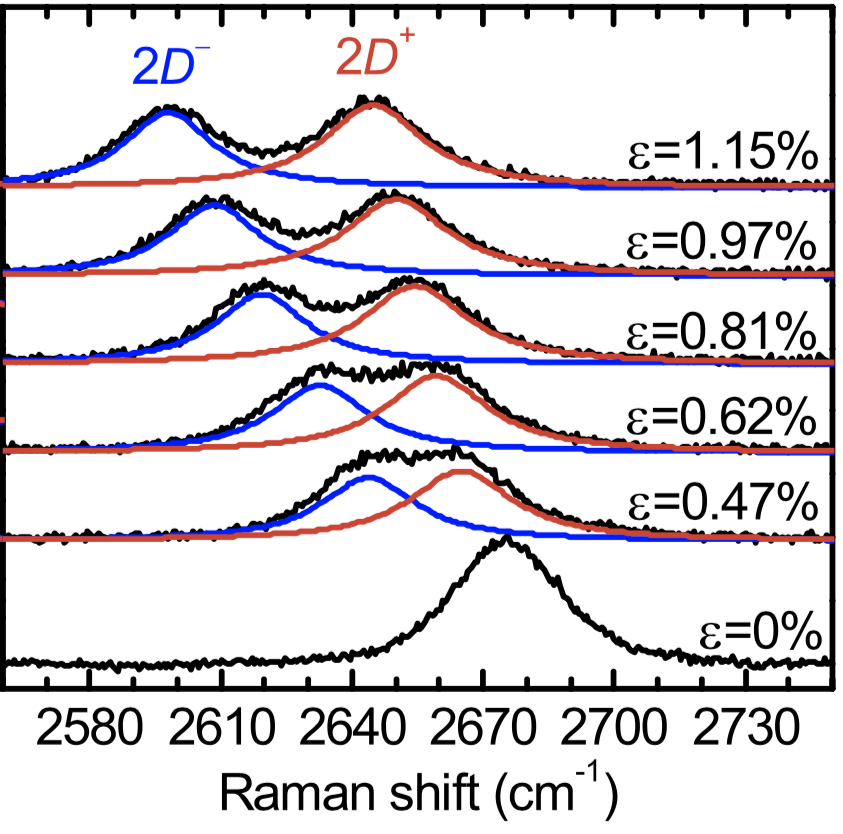}
	}
	\subfloat[]{
		\centering
		\includegraphics[scale=0.15]{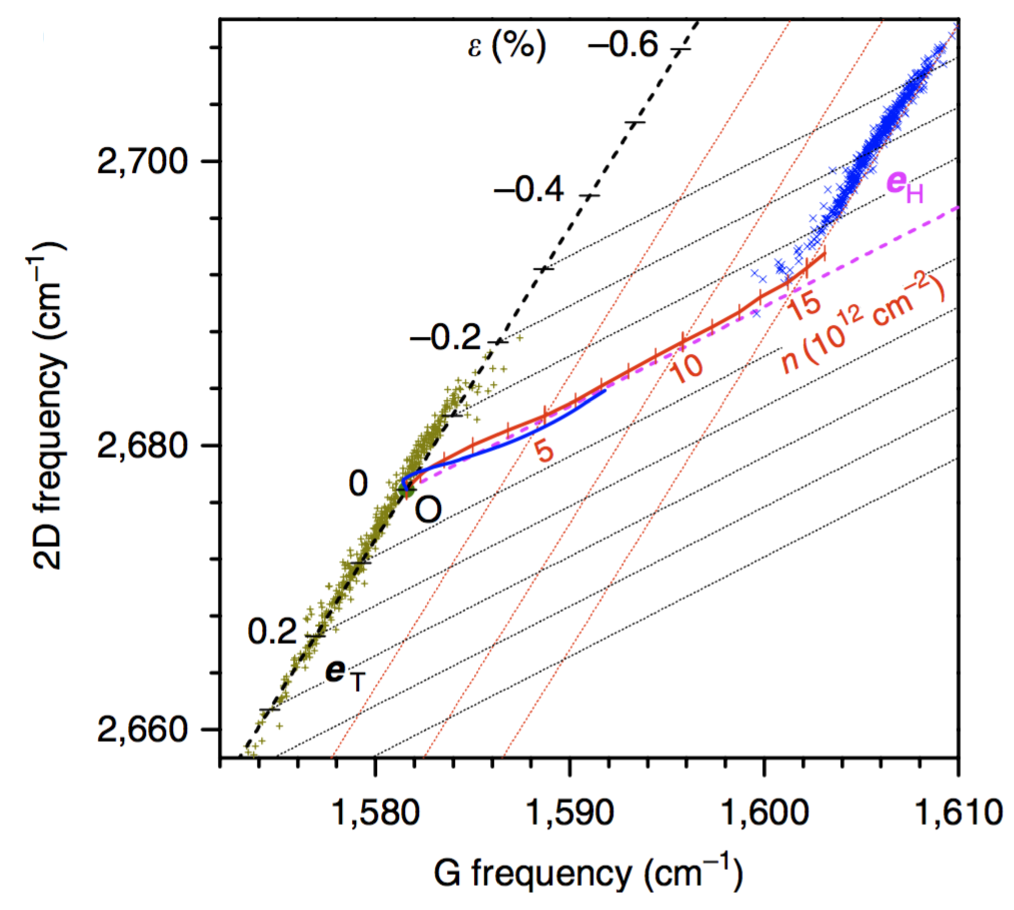}
	}
	\caption{
	{\bf Effects of strain on the Raman $G$ and $2D$ peaks.}
	(a,b) Influence of uniaxial strain on the Raman $G$ (a) and $2D$ (b) peaks.
	      Shown are experimentally recorded Raman spectra for increasing amounts of uniaxial strain, leading to a splitting of the $G$ and $2D$~peaks.
	      (Figures taken from Ref.~\onlinecite{mohiuddin2009} and Ref.~\onlinecite{yoon2011}, respectively.)
	(c) Illustration of the separation of doping and strain-induced shifts of the Raman $G$ and $2D$~peak positions as presented in Ref.~\onlinecite{lee2012}.
	    The two clusters of points represent measurements of the positions of the $G$ and $2D$~peaks on various spots of two different samples.
	    The dashed lines represent axes of slopes 0.7 and 2.2, corresponding to the linear relation between shifts of the $2D$ and $G$~peaks due to doping or strain, respectively.
	    	By projecting a point in the $(\omega_{G},\omega_{2D})$-plane onto these to axes, one can gain estimates for the amount of charge carrier doping and strain in the sample.
	    	The scaling of the axes is valid for hole doping and uniaxial strain, respectively.
	    (Figure taken from Ref.~\onlinecite{lee2012}.)
	}
	\label{fig:strain-influences}
\end{figure}

The $G$~peak is less affected by these changes than the two-phonon and/or defect-assisted peaks as the electronic structure only directly influences the intensity, i.e., the peak height.
The peak position and width, by contrast, are only determined by the frequency and width of the degenerate optical phonons at $\Gamma$.
The width of these phonons remains almost completely unaffected by strain since it is mostly determined by the decay of the phonon into electron-hole pairs.
The position, however, is rather sensitive to strain since it equal to the frequency of the phonon.
As mentioned before, strain leads to a change of the bond strength between the atoms, which in turn modifies the effective spring constant between the atoms and hence the phonon frequency.
In general, compressive strain leads to a hardening of the bonds as the charge density becomes higher between the atoms thus binding the atoms stronger together, which results in a higher phonon frequency and therefore a larger Raman shift.
Tensile strain, on the other hand, leads to the opposite effect and reduces the phonon frequency and the Raman shift.
For experimentally accessible strain values, the relation between Raman shift of the $G$~peak and strain is a linear one, with the constant of proportionality given by the \emph{Gr\"uneisen} parameter for the in-plane optical phonon at $\Gamma$.\\
\\
The two-phonon and/or defect-assisted peaks such as the $2D$~peak, however, undergo a larger change with strain.
Here, both the change in the phonon frequencies and the distortion of the electronic band structure play a role.
When the electronic band structure is distorted, the conditions for double resonance are changed.
When combined with the change of the phonon frequencies this leads to much larger shift of the two-phonon and/or defect-assisted peak positions with strain compared to that of the $G$~peak.\\
\\
Additionally, one has to distinguish between \emph{uniaxial} strain and \emph{biaxial} strain:

\emph{Uniaxial} strain means that strain is applied along a specific direction.
The fact that a specific direction is now singled out has a big influence on the various Raman peaks.
For the $G$~peak, the most important effect is the lifting of the degeneracy of the in-plane optical phonon frequencies at $\Gamma$.
In unstrained graphene, the two in-plane optical phonon branches are degenerate at $\Gamma$ because of rotation symmetry.
By singling out a specific direction, uniaxial strain breaks this rotation symmetry and as a result the two phonon branches split.
Since the Raman peak corresponding to one-phonon, defect-free processes is due to the emission of an in-plane optical phonon at $\Gamma$, the $G$~peak splits into two peaks corresponding to the two now non-degenerate phonon branches.
The difference between these two phonon frequencies is proportional to the applied strain and for low strain ($<$0.7\%~\cite{mohiuddin2009}), the peak splitting is too small compared to the width of the phonons to be observable.
Only for higher strain values does one record a clearly visible peak splitting, as seen in Fig.~\ref{fig:strain-influences}a.

For the two-phonon and/or defect-assisted peaks, uniaxial strain can also cause observable peak splitting.
The mechanism, however, is a different one than that for the $G$~peak.
The observed peak splitting of, for example, the $2D$~peak has its roots in a breaking of the threefold rotation symmetry of the resonance surface of the electron and phonon wave vectors.
Recall that the phonons mostly contributing to the peak are determined from resonance conditions derived in the previous section.
These resonance conditions can be fulfilled or approximately fulfilled by a whole set of wave vectors for the electrons and phonons, which however obey the lattice symmetries.
By breaking the threefold rotation symmetry of the lattice by applying uniaxial strain, the shape of the resonance surfaces in the Brillouin zone is distorted and this leads to different phonon wave vectors that satisfy them that correspond to phonon frequencies that are a bit separated from each other.
This leads to the observation of a splitting in the two-phonon and/or defect-assisted bands, as has been measured, for example, for the $2D$~peak~\cite{yoon2011} (compare Fig.~\ref{fig:strain-influences}b)
As in the case of the $G$~peak, the applied uniaxial strain has to be large enough to observe such a splitting with the threshold value being roughly in the same range as for the $G$~peak.

It should be noted that, for a low-quality graphene sheet and/or an unfortunate choice of substrate, the intrinsic width of the Raman peaks can be too large to allow the peak splitting to be observed.
Instead one will only measure a further broadening of the peaks (compare for instance Ref.~\onlinecite{mohiuddin2009}, where no $2D$~peak splitting was observed, and Ref.~\onlinecite{yoon2011}, where it was observed).

\emph{Biaxial} strain, on the other hand, does not lead to the splitting of the Raman peaks.
Due to the hexagonal lattice structure, biaxial strain leads mostly only to a uniform stretching or contraction of the unit cell and does not break the rotation symmetry in a major way.
Therefore, both mechanisms described in the previous paragraph do not apply in the case of biaxial strain and the Raman bands do not split.
However, the mechanism that leads to a shift of the peaks remains applicable.
As one would very naively expect, the change of the peak positions for a given amount of strain is indeed much larger for biaxial strain than for uniaxial strain~ \cite{lee2012}.\\
\\
Finally, we want to point out that while strain and doping both lead to a shift of the positions of the various peaks, it has recently been demonstrated that it is possible to disentangle these two sources of peak shifts~\cite{lee2012}.
By focusing on the positions of $G$ and $2D$~peaks, one can exploit the fact that the two peak shifts are linearly correlated, both for doping and strain as the source of the peak shifts.
The ratio of the two shifts, however, depends on the source of the shifts:
For hole doping, the ratio of the $2D$ and $G$~peak shifts is roughly $\Delta\omega_{2D}/\Delta\omega_G|_{\text{hole dop.}} \sim 0.7$.
For electron doping, the slope is much slower and deviates soon from the linear trend.
For strain, the $2D$~peak, as being related to a two-phonon Raman process, is subject to much larger changes of the position compared to the $G$~peak, with the experimentally extracted ratio of the peak shifts given by $\Delta\omega_{2D}/\Delta\omega_G|_{\text{strain}} \sim 2.2$.
Under the assumption that the doping- and strain-induced shift are simply additive and not correlated, one can project the coordinates of a point in the $\omega_G$-$\omega_{2D}$ plane onto the axes of slopes 0.7 and 2.2 and separate the doping- and strain-induced shifts (compare Fig.~\ref{fig:strain-influences}c), whereupon both can be converted to a value for the charge carrier concentration or the strain, respectively, using theoretical predictions.

\section{Conclusions}

This chapter has given a glimpse at the rich variety of phenomena that occur in Raman spectroscopy of graphene.
Due to the linear band structure of graphene, there are always electron-hole pairs in resonance with the incoming/outgoing laser light.
This leads to the relatively large Raman cross section and the strong sensitivity of the Raman spectrum on the sample parameters.
A full quantitative explanation of Raman peak positions, peak widths, and intensities requires the somewhat lengthy formalism of time-dependent perturbation theory.
However, we have seen that kinematic considerations and a careful look at the matrix elements and energy denominators allow in many cases an intuitive understanding of the Raman spectrum without explicit calculations.

The $G$~line position is given by the $E_{2g}$ phonon frequency at $\Gamma$ and thus it does not disperse as a function of the laser energy.
Its position and width are sensitive measures for doping and strain.
At the same time it remains largely unaffected by defects and by the number of layers.
The $2D$~line, by contrast, disperses as a function of the laser wave length.
Its splitting into subpeaks (sometimes only seen as an increased broadening of the line) is, in particular, used as a convenient measure for the number of layers in the sample.
Its position is also affected by strain, but only moderately by doping.
The intensity of defect related peaks is a direct measure of the defect concentration in the sample.

One of the main challenges for quantitative calculations is still the absolute position and the width of the $2D$~line (and thus also of the $D$~line).
The difficulties are twofold:
(i) a fully {\it ab initio} calculation of the highest optical phonon branch around $K$ is still missing (only the linear slope at $K$ has been determined quantitatively using the $GW$~approximation).
(ii) The width is a result of both the effect of k-space integration and also depends on the various broadenings of intermediate electron and hole states as well as on the phonon frequency.

For the time being, {\it ab initio} calculations of the Raman spectrum of graphene (such as, for example, the calculation presented in Ref.~\onlinecite{herziger2014}) still rely on the input of phonon and electron lifetimes as semi-empirical parameters.
All the lifetimes (due to electron-electron, electron-phonon, and phonon-phonon interaction) can, in principle, be calculated, but calculations remain quite expensive in terms of computing time.
With the increasing available computing power and recent progress in code development for lifetime calculations, we are optimistic that a fully {\it ab initio} calculation of the Raman spectrum of graphene can be achieved in the near future.

\begin{acknowledgments}

We acknowledge support by the National Research Fund, Luxembourg (projects FNR10196198 and FNR7490149).
Furthermore, we acknowledge stimulating discussions with Franscesco Mauri, Michele Lazzeri, and Matteo Calandra.

\end{acknowledgments}



%


\end{document}